\documentclass[onecolumn,superscriptaddress,
 amsmath,amssymb,
 aps,
 pre,longbibliography,
]{revtex4}

\usepackage[utf8]{inputenc}
\usepackage{amsmath}
\usepackage[T1]{fontenc}
\usepackage{amsmath,amsfonts,amssymb,amsthm,epsfig,epstopdf,url,array}
\usepackage{bm}
\usepackage{graphicx}
\usepackage{mathtools}
\usepackage{physics}
\usepackage{comment}
\usepackage{color}

\usepackage{dcolumn}
\usepackage{bm}
\usepackage{graphicx}
\usepackage{comment}

\newcommand{\bea}{\begin{eqnarray}}

\newcommand{\eea}{\end{eqnarray}}

\usepackage{color}

\newcommand{\blue}{\textcolor{black}}

\let\oldref\ref
\renewcommand{\ref}[1]{(\oldref{#1})}

\usepackage[utf8]{inputenc}
\usepackage{amsmath}
\usepackage[T1]{fontenc}
\usepackage{amsmath,amsfonts,amssymb,amsthm,epsfig,epstopdf,url,array}
\usepackage{bm}
\usepackage{graphicx}
\usepackage{mathtools}
\usepackage{physics}
\usepackage{comment}
\usepackage{color}

\usepackage{dcolumn}
\usepackage{bm}
\usepackage{graphicx}
\usepackage{comment}

\usepackage{color}

\let\oldref\ref
\renewcommand{\ref}[1]{(\oldref{#1})}

\usepackage[utf8]{inputenc}      
\usepackage[T1]{fontenc}         
\usepackage[english]{babel}      
\usepackage{amsmath,amssymb}     
\usepackage{graphicx}            
\usepackage{float}               
\usepackage{booktabs}            
\usepackage[colorlinks=true,linkcolor=blue,citecolor=blue,urlcolor=blue]{hyperref}
\usepackage{enumitem} 

\begin{document}

\title{Dynamically emergent correlations in Brownian particles subject to simultaneous non-Poissonian resetting protocols}

\author{Gabriele de Mauro}
\affiliation{LPTMS, CNRS, Univ.  Paris-Sud,  Universit\'e Paris-Saclay,  91405 Orsay,  France}
\author{Marco Biroli}
\affiliation{LPTMS, CNRS, Univ.  Paris-Sud,  Universit\'e Paris-Saclay,  91405 Orsay,  France}
\author{Satya N. Majumdar}
\affiliation{LPTMS, CNRS, Univ.  Paris-Sud,  Universit\'e Paris-Saclay,  91405 Orsay,  France}
\author{Gr\'egory Schehr}
\affiliation{Sorbonne Universit\'e, Laboratoire de Physique Th\'eorique et Hautes Energies, CNRS UMR 7589, 4 Place Jussieu, 75252 Paris Cedex 05, France}

\begin{abstract}
We consider a one-dimensional gas of \( N \) independent Brownian particles subject to simultaneous stochastic resetting, with inter-reset times drawn from a general waiting-time distribution \( \psi(\tau) \). This includes the well-known Poissonian case, where \( \psi(\tau)=re^{-r\tau} \), and extends to more general classes of resetting, such as heavy-tailed and bounded distributions. We show that the simultaneous resetting generates correlations between particles dynamically. These correlations grow with time and eventually drive the system into a strongly correlated non-equilibrium stationary state (NESS). Exploiting the renewal structure of the resetting dynamics, we derive explicit analytical expressions for the joint distribution of the positions of the particles in the NESS. We show that the NESS has a conditionally independent and identically distributed (CIID) structure that enables us to compute various physical observables exactly for arbitrary $\psi(\tau)$. These observables include the average density, 
extreme value and order statistics, the spacing distribution between consecutive particles and the full counting statistics, i.e., the distribution of the number of particles in a given interval centered at the origin. We discuss the universal features of the large $N$ scaling behaviors of these observables for different choices of the resetting protocol \( \psi(\tau) \).  Our results provide an interesting example of a stochastic control whereby, by tuning the inter-reset distribution $\psi(\tau)$, one can generate a class of tunable, and yet solvable, strongly correlated NESS in a many-body system. 
\end{abstract}

\maketitle

\section{Introduction}\label{sec:introduction}

It is quite common in everyday life for certain processes to be interrupted at random times and then restarted from a fixed initial condition. For example, when looking for misplaced keys at home, people often return repeatedly to a known starting point (e.g., the entryway or a specific room) to restart their search if the keys are not found within a reasonable timeframe. Similarly, during a visual search in a crowded place, our gaze tends to return to a fixed starting position before resuming the search when the target is not immediately found.

A prototypical example of such processes, known as stochastic resetting, was introduced in \cite{DiffusionwithStochasticResetting}, where a diffusive particle is reset to its initial position at a constant rate 
$r$. Since then, stochastic resetting has been studied in a wide variety of contexts (for recent reviews see \cite{SR_review,pal2022inspection,gupta2022stochastic}). In particular, it has found practical applications in many search problems \cite{Diffusionwithoptimalresetting,Optimalresettingstrategiesforsearchprocessesinheterogeneousenvironments,tal2025smart,Searchprocesseswithstochasticresettingandmultipletargets,Activemotioncanbebeneficialfortargetsearchwithresettinginathermalenvironment,OptimalStochasticRestart,FirstPassageunderRestart,RandomSearchwithResetting,Searchwithhomereturns,DiffusiveSearchwithspatiallydependent,BrownianBridges,Criticalnumberofwalkers,time-dependentresetting,hartmann2025diffusion,evans2014diffusion,evans2018effects,bodrova2020resetting,biswas2025target,FirstOrderTransition...,del2025proxitaxis},  where, by tuning the resetting rate to an optimal value, it is possible to improve the mean time to find the target under certain conditions. Resetting has also been employed in computer simulations~\cite{StochasticResettingforEnhancedSampling} to improve sampling efficiency in molecular dynamics algorithms, as it helps the system escape ``bad'' local minima. More generally, incorporating a resetting mechanism into stochastic algorithms may help them avoid getting stuck before completing their execution \cite{algor_1,algor_2,algor_3,algor_4,Algor_5}.
In biology, stochastic resetting is used to model RNA synthesis, where transcription is stochastically interrupted via backtracking \cite{Backtrack,backtrack_2}. 

Resetting the dynamics of a physical system has a profound effect on its late time behavior: the resetting moves break detailed balance and typically drive the system into a non-equilibrium stationary state (NESS). In general, non-equilibrium systems remain poorly understood, and stochastic resetting offers a valuable framework for constructing broad classes of such systems where the stationary state can be computed explicitly. This makes it a powerful tool for gaining deeper insights into the behavior of systems far from equilibrium.
The implications of this mechanism have been extensively explored in single-particle systems, such as L\'evy walks, L\'evy flights, and run-and-tumble particles \cite{Runandtumbleparticleunderresetting:arenewalapproach,run_and_t_2,FirstOrderTransition...,Continuous-timerandomwalksandLevywalks}. Analytical predictions from resetting systems have also been verified experimentally using colloidal particles in harmonic traps \cite{Besga_2020,Experimental_3,Experimental_1,Many-BodyColloidalDynamics...,biroli2025exp}. Resetting has also been studied in a variety of quantum systems~\cite{Quantum_1,Quantum_2,Quantum_3,Quantum_4}, with various applications, including the random measurement and detection of quantum states~\cite{kessler2021first,perfetto2021designing,das2022quantum,magoni2022emergent,kulkarni2023first,kulkarni2023generating,yin2023restart,chatterjee2024quest,yin2024instability,yin2025restart,roy2025causality}. Moreover, the \emph{single-particle} problem has been generalized to arbitrary waiting time distributions \( \psi(\tau) \) between resets, including power-law distributions \cite{Powerlawreset,Non-equilibriumsteadystatesof...,bodrova2020continuous} and time-dependent resetting rates \( r(t) \) \cite{time-dependentresetting}.

Finally, resetting was shown to profoundly change the behavior of many body systems~\cite{FluctuatingInterfaces...,Symmetricexclusionprocess..,Isingmodel..,MCPL2022,gas_bosons,Exact_Extreme,Biroli_Harmonic_trap,biroli2025resetting}. In particular, if all the degrees of freedom are reset simultaneously the resetting event dynamically generates correlations between all the degrees of freedom. A recent example consists of a gas of $N$ independent Brownian motions on a line that are reset simultaneously to the origin with a rate $r$ \cite{gas_bosons}. In this system, even though the particles are noninteracting between two successive resets, the simultaneous resetting induces strong correlations between these particles that grow dynamically. Eventually the system approaches a NESS where the particles get strongly correlated (with attractive all-to-all correlations). Remarkably, despite the presence of strong correlations and the non-equilibrium nature of the steady state, many observables of physical interest remain analytically tractable \cite{gas_bosons,Exact_Extreme}. These observables include
\begin{itemize}
    \item {the average density of particles};
    \item {the order statistics}, i.e., the statistics of the ordered positions of the particles. For example, the first maximum, second maximum, etc, when the particles are increasingly ordered from the right. This includes the extreme value statistics (EVS) as a special case, where one studies the statistics of the maximum (the rightmost position) or the minimum (the leftmost position);
 \item {the gap statistics}, i.e., the distribution of the gap between the positons of two consecutive particles;
 \item {the full counting statistics (FCS)}, i.e., the distribution of the number of particles $N_L$ that lie in the interval \([-L, L]\) centered at the origin.
\end{itemize}
The possibility to analytically calculate all these quantities is a very rare feature in a typical non-equilibrium system.

 Among all these observables, the study of EVS is of particular interest. While the classical theory of EVS classifies the distribution of the maximum of independent and identically distributed (IID) variables into Gumbel, Fr\'echet, and Weibull \cite{AFirstCourseinOrderStatistics,OrderStatistics} universality classes, much less is known in the presence of strong correlations. Some generalisations have been obtained for weakly correlated variables \cite{Extremevaluestatisticsofcorrelatedrandomvariables:Apedagogicalreview,StatisticsofExtremesandRecordsinRandomSequence}, or non identically distributed random variables \cite{Weissman_1988,krug2007records, lacroix2019rotating}, but for strongly correlated variables there are no general results and only a few cases have been solved \cite{Extremevaluestatisticsofcorrelatedrandomvariables:Apedagogicalreview,Exactrecordandorderstatisticsofrandomwalksviafirst-passageideas}.
Resetting provides a rare example where such statistics can still be computed exactly, even in the correlated regime.

So far, many-body systems with resetting have been studied only for Poissonian resetting protocol, where $\psi(\tau) = r\,e^{-r\tau}$, with $r$ denoting the resetting rate. In contrast, single particle systems have been studied for general non-Poissonian resetting protocols $\psi(\tau)$. The purpose of this paper is to study many-body systems subject to a generic non-Poissonian resetting protocol $\psi(\tau)$, thus generalizing the Poissonian protocol to many-body systems. We focus on a one-dimensional gas of \( N \) independent Brownian particles, where all particles are reset simultaneously to their initial positions. The intervals between resets are drawn from a general waiting time distribution \( \psi(\tau) \). We first review the known results for the Poissonian case, where \( \psi(\tau) = r e^{-r\tau} \), and then analyze how the NESS changes under two specific non-exponential distributions $\psi(\tau)$: a power-law and a bounded distribution. This analysis allows us to characterize the NESS more broadly in terms of the resetting protocol \( \psi(\tau) \), identifying which features are universal and which depend on the specific form of \( \psi(\tau) \). All the observables introduced earlier are analyzed within this general setting.

Let us briefly recall the main features of the NESS for $N$ independent Brownian motions subjected to simultaneous resetting via the Poissonian protocol $\psi(\tau) = r\,e^{-r\tau}$. The particles all start from the origin and reset simultaneously to the origin. Between two consecutive resets, the 
positions of the particles $x_i$'s evolve via free diffusion. The joint position distribution is then governed by the free Gaussian propagator
\begin{equation}\label{eq:propagator_factorizable}
    \mathcal{G}_0(\vec{x},t) = \prod_{i=1}^{N}\frac{1}{\sqrt{4\pi D t}}\exp\left(-\frac{x_i^2}{4Dt}\right) \;,
\end{equation}
where \( D \) is the diffusion constant and $\vec{x} \equiv \{x_1, x_2, \cdots, x_N\}$ denotes the positions of the particles. The subscript $0$ indicates that it corresponds to the joint probability density function (JPDF) of the positions when no resetting is present (\( r=0 \)). How does resetting change this propagator? The JPDF \( \mathcal{P}_r(\vec{x},t) \), representing the probability density of finding particles at positions \( \vec{x} \) at time \( t \) under a constant resetting rate \( r \), can be derived explicitly by exploiting the renewal structure of the process. This renewal approach will also prove useful when generalizing the model to a non-exponential \( \psi(\tau) \).
The key observation is that whenever a reset occurs, the system restarts from the initial condition, rendering its previous history irrelevant.  
As a result, the state of the system at time \( t \) depends solely on the time elapsed since the \emph{last} reset.    
The JPDF can thus be expressed as the sum of two distinct contributions:
\begin{equation}\label{eq:last_renew_poissonian}
    \mathcal{P}_r(\vec{x},t) = e^{-rt}\mathcal{G}_0(\vec{x},t) + r\int_0^t d\tau\, e^{-r\tau}\, \mathcal{G}_0(\vec{x},\tau) \,.
\end{equation}

\begin{figure}    \centering
    \begin{minipage}{0.49\textwidth}
        \centering
        \includegraphics[width=\linewidth]{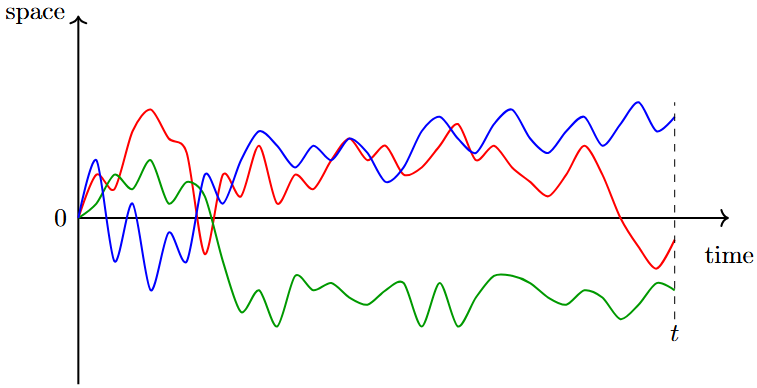}
    \end{minipage}
    \hfill
    \begin{minipage}{0.49\textwidth}
        \centering
        \includegraphics[width=\linewidth]{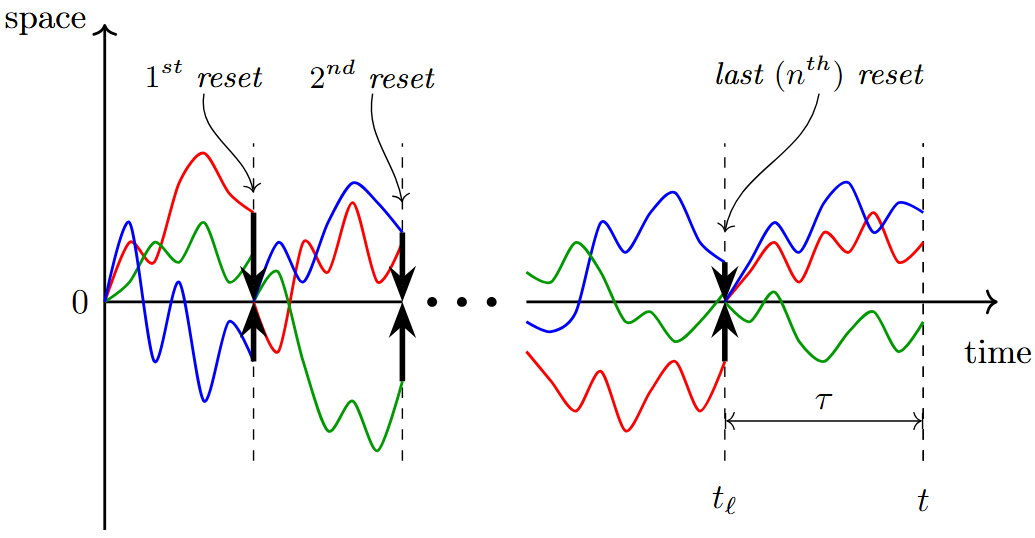}
    \end{minipage}
    \caption{\textbf{Schematic illustration of the process.} 
    The left panel shows the sketch of a trajectory where no reset occurs up to time~$t$: the particles evolve purely by free diffusion.
    The right panel shows the sketch of a trajectory that undergoes at least one reset before time~$t$. The last reset occurs at time~$t_\ell$, and there are $n-1$ resets before it, with $n \geq 1$. We explicitly show the first, the second, and the last $(n^{th})$ reset.
    Since each reset completely deletes the system's past, the state of the system at time $t$ depends only on the time interval $\tau = t - t_{\ell}$ since the last reset.}
    \label{fig:illustrazione_processo}
\end{figure}
The first term corresponds to the trajectories where no resetting has occurred up to time $t$, which happens with probability $e^{-rt}$. In this case, the JPDF is simply
that of the process without resetting $\mathcal{G}_0(\vec{x},t)$. The second term considers instead the trajectories where the last resetting has occurred at time $t-\tau$, and then the particles propagate freely in the subsequent interval $[t-\tau,t]$. An illustration of the process can be seen in Fig. \ref{fig:illustrazione_processo}. In this latter case, the probability of a reset at time $t-\tau$ and then no reset up to $t$ is simply \blue{$re^{-r\tau}$}.
In the limit $t \to \infty$ the first term in \eqref{eq:last_renew_poissonian} drops out and the stationary state is be given by
\begin{equation}\label{eq:SS_poissonian}
    \mathcal{P}^*_r(\vec{x}) = r\int_0^{\infty} d\tau\, e^{-r\tau}\, \prod_{i=1}^{N} \frac{1}{\sqrt{4\pi D \tau}} \exp\!\left(-\frac{x_i^2}{4D\tau}\right) \,.
\end{equation}
From this expression, we can clearly see that the JPDF in the NESS does not factorise, indicating that the positions of the particles are clearly correlated in the stationary state. However, the JPDF has a special conditionally independent and identically distributed (CIID) structure, namely if one conditions on the random variable $\tau$, drawn from $r\,e^{-r\tau}$, then the variables $x_i$'s are independent and identically distributed (IID), each drawn from the conditional PDF $p_0(x\vert \tau) = e^{-x^2/(4 D\tau)}/\sqrt{4 \pi D\,\tau}$, parametrized by $\tau$. Therefore the statistics of any observable in the NESS can be computed in two steps: (i) compute it for the IID variables $x_i$'s conditioned on fixed $\tau$ (this is akin to calculating these observables for an ``ideal'' Brownian gas parametrized by $\tau$) and (ii) average over $\tau$ drawn from $r\,e^{-r\tau}$. The step (i) greatly simplifies the problem since most of the observables mentioned before can be easily computed for IID random variables \cite{Extremevaluestatisticsofcorrelatedrandomvariables:Apedagogicalreview,StatisticsofExtremesandRecordsinRandomSequence}. This CIID structure in the NESS is not limited to the resetting Brownian ideal gas described above. Indeed, it has recently been found in a number of classical and quantum models~\cite{gas_bosons,Exact_Extreme,sabhapandit2024noninteracting,kulkarni2025dynamically,dean2025exact,mesquita2025dynamically,biroli2025exp,mesquita2025dynamicallyb}.

The goal of this paper is to show that a similar CIID structure in the NESS also holds for a generic non-Poissonian resetting protocol $\psi(\tau)$, not necessarily a pure exponential as in the Poissonian case. This fact allows us to compute several physical observables exactly in the NESS for arbitrary resetting protocols $\psi(\tau)$, thus generalizing the results known previously only for the Poissonian protocol. The rest of the paper is organized as follows. 
In Section \ref{sec:model}, we define the model precisely with a non-Poissonian resetting protocol $\psi(\tau)$ and derive the exact JPDF at any time $t$ and discuss the conditions under which a stationary state exists. We then demonstrate how the correlations grow with time, saturating finally to their stationary value, when such a stationary state exists. In Section \ref{sec:NESS} we derive the stationary JPDF when it exists and discuss the different macroscopic and microscopic observables in the NESS. These include the average density, the order statistics, the spacing distribution and the FCS. In Section \ref{sec:three_examples}, we present explicit results for three different resetting protocols: the Poissonian, the power-law and the bounded $\psi(\tau)$. In Section \ref{sec:risultati}, we present the results for general $\psi(\tau)$ and discuss the universal features of the large $N$ scaling behavior of different observables. Finally, Section \ref{sec:concl} contains a summary and outlook. Some details of the computations are relegated to the Appendices \ref{app:iidvar} and \ref{app:appendix_calcoli}.

\section{The model of diffusing particles with a generalized non-Poissonian resetting dynamics}\label{sec:model}

We start by recalling the Poissonian resetting dynamics with rate $r$ of $N$ independent Brownian particles on a line, all starting at the origin $\vec{x}_0 = \vec{0}$~\cite{gas_bosons}. In a small time interval $dt$, the positions of the particles are updated via 
\bea \label{dis_Langevin}
x_i(t+\Delta t) = 
\begin{cases}
& 0 \; \; \hspace*{3.cm} {\rm with \; prob.}\; r \, \Delta t \;, \;\hspace*{1.cm}\quad {\rm for \; all} \quad \; i=1, \cdots N \;,\\
& \\
& x_i(t) + \sqrt{2D}\,\eta_i(t)\, dt \quad {\rm with \; prob.}\; 1-r \, \Delta t\;, \quad \;\quad {\rm for \; all} \quad \; i=1, \cdots N \;,
\end{cases}
\eea
where $\eta_i(t)$ are independent Gaussian white noises satisfying $\langle \eta_i(t) \eta_j(t') \rangle = \delta_{ij} \delta(t - t')$. The first term corresponds to the simultaneous resetting to the origin of all the $N$ particles, while the second term describes local diffusion with a common diffusion constant $D$. In terms of trajectories, this dynamics (\ref{dis_Langevin}) is equivalent to the following: starting at the origin, we let $N$ particles propagate independently up to a random time $\tau$ distributed via an exponential distribution $\psi(\tau)= r e^{-r \tau}$. Following this free propagation of duration $\tau$, the positions of the particles are reset instantaneously (and simultaneously) to the origin. Following the reset, again the particles evolve independently up to another random time $\tau$, drawn independently from $\psi(\tau)= r e^{-r \tau}$ and then get simultaneously reset to the origin. The process continues ad infinitum. 

It is then straightforward to generalize this Poissonian resetting model to the more general non-Poissonian case where $\psi(\tau)$ is an arbitrary normalisable distribution, i.e., 
\bea \label{norm_psi}
\int_0^\infty \psi(\tau) \, d\tau = 1 \;. 
\eea
In this case, while one can not write down a memoryless Langevin equation as in the Poissonian case in Eq. (\ref{dis_Langevin}), one can easily define the model in terms of the dynamics of the trajectories, as illustrated in the right panel of Fig. \ref{fig:illustrazione_processo}. We start again with $N$ Brownian particles at the origin at $t=0$ and propagate them freely and independently up to a random time $\tau$ drawn from $\psi(\tau)$. At the end of this interval, the particles are instantaneously and simultaneously reset to the origin, as in the Poissonian case. After this, the process again repeats via alternating free Brownian propagation during a random interval $\tau$ drawn from $\psi(\tau)$ and the instantaneous and simultaneous resetting. Below, using a renewal approach, we first derive exactly the JPDF of the positions of the particles at any time $t$.

\subsection{Derivation of the exact joint distribution at any time $t$ via a renewal approach}

The position distribution of a single particle subject to non-Poissonian resetting has been investigated using several different approaches, as summarized in \cite{SR_review}. In particular, the case of power-law resetting times is studied in \cite{Powerlawreset}, while the effect of a time-dependent resetting rate \( r(t) \) is discussed in \cite{time-dependentresetting}. More generally, \cite{Non-equilibriumsteadystatesof...} introduces a generalized master equation that captures arbitrary waiting time distributions between resets.

In this work, we provide a derivation for the position distribution of $N$ particles at any time $t$ for the non-Poissonian resetting by adapting a renewal approach that was used before for the Poissonian case \cite{gas_bosons}. This method classifies all trajectories up to time $t$ that contribute to the joint probability density function (JPDF) $\mathcal{P}(\vec{x}, t)$ into two categories:
\begin{enumerate}
    \item Trajectories that have not experienced any reset up to time $t$;
    \item Trajectories that have experienced at least one reset before time $t$.
\end{enumerate}

These two types of trajectories are illustrated in Figure~\ref{fig:illustrazione_processo}.  
To compute their contribution to the total JPDF, we introduce the function $\Psi(t) = \int_t^\infty \psi(\tau')\, d\tau'$, which gives the probability that no reset has occurred up to time $t$. For trajectories of the first type, the system evolves via free diffusion over the interval $[0, t]$, contributing $\Psi(t)\, \mathcal{G}_0(\vec{x}, t)$ to the JPDF, where $\mathcal{G}_0$ is the free propagator defined in Eq.~\eqref{eq:propagator_factorizable}.

We now consider trajectories that have experienced at least one reset before time $t$. Due to the renewal nature of the process, the system's statistical properties at time $t$ depend only on the time since the \textit{last} reset, not on the entire history. Let $t_\ell \in [0, t]$ denote the time of the last reset. The key object is $\Upsilon(t_\ell)$, the probability density that the last reset occurred exactly at time $t_\ell$, regardless of how many previous resets happened before. Given $\Upsilon(t_\ell)$, the contribution to the JPDF from such trajectories is
\begin{equation}
\int_0^t dt_\ell\, \Upsilon(t_\ell)\, \Psi(t - t_\ell)\, \mathcal{G}_0(\vec{x}, t - t_\ell).
\end{equation}
In other words, the last reset occurs at time $t_\ell$ with probability $\Upsilon(t_\ell)\, dt_\ell$. From that point until time $t$, no further reset occurs, which happens with probability $\Psi(t - t_\ell)$. During this interval, the particles evolve freely according to the propagator $\mathcal{G}_0(\vec{x}, t - t_\ell)$. Since $t_\ell$ can take any value in $(0, t)$, we integrate over it. Using $ \mathcal{G}_0(\vec{x}, t)$ from Eq. (\ref{eq:propagator_factorizable}), one thus obtains the JPDF at any time $t$ for a general resetting protocol $\psi(\tau)$
\begin{equation}\label{eq:last_renewal}
\mathcal{P}(\vec{x}, t) = \Psi(t)\, \prod_{i=1}^{N}\frac{1}{\sqrt{4\pi D t}}e^{-\frac{x_i^2}{4Dt}}
+ \int_0^t dt_\ell\, \Upsilon(t_\ell)\, \Psi(t - t_\ell)\, \prod_{i=1}^{N}\frac{1}{\sqrt{4\pi D (t-t_\ell)}}e^{-\frac{x_i^2}{4D(t-t_\ell)}}\;.
\end{equation}

Thus the only quantity that we need to determine is $\Upsilon(t_\ell)$. In fact, it can be expressed in terms of the interval distribution $\psi(\tau)$ as follows. We can express $\Upsilon(t_\ell)$ as a sum
\begin{equation}\label{eq:upsilon}
\Upsilon(t_\ell) = \sum_{n=1}^\infty \upsilon_n(t_\ell) \,,
\end{equation}
where $\upsilon_n(t_\ell)$ is the probability density that exactly the $n^{th}$ reset occurs at time $t_\ell$. Since any number of resets can happen before the last one, we sum over all $n \geq 1$. Next we notice that the functions $\upsilon_n(t)$ satisfy the recursion relation
\bea \label{upsilon_n}
\upsilon_n(t) = \int_0^t d\tau\, \psi(t - \tau)\, \upsilon_{n-1}(\tau), \quad \text{with} \quad \upsilon_1(t) = \psi(t) \;.
\eea
We can exploit the convolution structure in Eq. (\ref{upsilon_n}) by taking its Laplace transform with respect to (w.r.t.) $t$. We denote the Laplace transforms by
\bea \label{def_LT}
\tilde \psi(s) = \int_0^\infty e^{-st}\, \psi(t)\, dt \quad {\rm and} \quad \tilde{\upsilon}_n(s) = \int_0^\infty e^{-st}\, \upsilon_n(t)\, dt \;.
\eea
It then follows from Eq. (\ref{upsilon_n}) that $\tilde{\upsilon}_n(s) = [\tilde{\psi}(s)]^n$.
Summing over all $n \geq 1$, we obtain
\bea \label{def_upsilon}
\tilde{\Upsilon}(s) = \sum_{n=1}^\infty [\tilde{\psi}(s)]^n = \frac{\tilde{\psi}(s)}{1 - \tilde{\psi}(s)}.
\eea
It is also convenient to introduce the Laplace transform of the propagator w.r.t. $t$, i.e.,
\bea \label{def_LT}
\tilde{\mathcal{P}}(\vec{x},s) = \int_0^\infty \mathcal{P}(\vec{x}, t)\, e^{-st}\, dt  \;.
\eea  
Taking the Laplace transform of equation~\eqref{eq:last_renewal} w.r.t. $t$ and using the convolution form of the second term
on the right-hand side (RHS) of~\eqref{eq:last_renewal} gives
\bea \label{LT_2}
\tilde{\mathcal{P}}(\vec{x},s) = \left(1 +\tilde \Upsilon(s)\right) \int_0^{\infty} d\tau e^{-s\tau} \Psi(\tau) \mathcal{G}_0(\vec{x}, \tau) =  \frac{1}{1 - \tilde{\psi}(s)} \int_0^{\infty} d\tau\, e^{-s\tau}\, \Psi(\tau)\, \mathcal{G}_0(\vec{x}, \tau) \;,
\eea
where we used the expression of \blue{$\tilde{\Upsilon}(s)$ } in Eq. (\ref{def_upsilon}). Recalling that $\psi(t) = -\frac{d\Psi(t)}{dt}$, the Laplace transform of $\Psi(t)$ is given by
\bea \label{res_Psi}
\tilde{\Psi}(s) = \frac{1 - \tilde{\psi}(s)}{s}.
\eea
Inserting this into equation~\eqref{LT_2} leads to the final expression
\begin{equation}\label{eq:Stat_state_laplace}
\tilde{\mathcal{P}}(\vec{x},s) = \frac{1}{s\, \tilde{\Psi}(s)} \int_0^{\infty} d\tau\, e^{-s\tau}\, \Psi(\tau)\, \prod_{i=1}^{N}\frac{1}{\sqrt{4\pi D \tau}}e^{-\frac{x_i^2}{4D \tau}} \;.
\end{equation}
Thus, knowing the interval distribution $\psi(\tau)$ between successive resettings, along with the explicit propagator $\mathcal{G}_0(\vec{x}, t)$ given in Eq. (\ref{eq:propagator_factorizable}), completely determines the joint distribution of the position of the particles $\mathcal{P}(\vec{x}, t)$ at any time~$t$. This is the main new result in this section.  Note that this result (\ref{eq:Stat_state_laplace}) can alternately be derived very simply by applying 
the renewal method with a first resetting approach \cite{DiffusionwithStochasticResetting}, but here, for completeness, we have included a derivation based on the last resetting approach.

\blue{For the Poissonian case, where $\psi(\tau) = r\,e^{-r\tau}$ and $\Psi(t) = \int_t^\infty \psi(\tau) \, d\tau= e^{-rt}$, Eq. (\ref{eq:Stat_state_laplace}) gives the exact expression~\cite{gas_bosons}
\bea \label{laplace_Pois}
\tilde{\mathcal{P}}(\vec{x},s) = \frac{1}{(2 \pi)^{N/2}} \frac{\| \vec{x} \|^{1-N/2}}{s\, D^{(N+2)/4}} (r+s)^{(N+2)/4} K_{(N-2)/2}\left( \sqrt{\frac{r+s}{D}} \| \vec{x} \|
 \right) \;,
\eea
where $ \| \vec{x} \|= \sqrt{x_1^2+...,x_N^2}$ and $K_\nu(z)$ is the modified Bessel function of the second kind. For $N=1$ it reduces to the well known expression \cite{SR_review}
\bea \label{expr_N1}
\tilde{\mathcal{P}}(x,s) = \frac{1}{2s} \sqrt{\frac{r+s}{D}}   \, e^{- \sqrt{\frac{r+s}{D}} |x|} \;.
\eea}

\subsection{Dynamically growing correlations}

From the expression of the joint distribution in Eq. (\ref{eq:last_renewal}), one sees clearly that it does not factorize at any time $t$, indicating the presence of a nonzero correlation between the particles. To characterise the correlations, it is natural to study the correlation function
\bea \label{def_C1}
{\cal C}_1(t) = \langle x_i(t) x_j(t) \rangle - \langle x_i(t)  \rangle \langle  x_j(t) \rangle \quad, \quad i \neq j \;,
\eea
where the angular brackets $\langle \cdots \rangle$ denote an average w.r.t. the JPDF in Eq. (\ref{eq:last_renewal}). We note immediately that this JPDF is invariant under the flip $x_i \to -x_i$ and consequently $C_1(t) = 0$ for any time $t$. Hence to detect the correlations, we need to investigate higher order correlation functions. The simplest nonzero correlator turns out to be \cite{gas_bosons} 
\bea \label{def_C2}
{\cal C}_2(t) = \langle x^2_i(t) x^2_j(t) \rangle - \langle x^2_i(t)  \rangle \langle  x^2_j(t) \rangle \quad, \quad i \neq j \;. 
\eea
This can be computed easily from Eq. (\ref{eq:last_renewal}) since, for $i \neq j$, it involves computing the second moment of a Gaussian distribution. We get
\bea \label{expr_C2}
{\cal C}_2(t) = 4 D^2 \left[ t^2 \Psi(t) + \int_0^t d\tau \Upsilon(t-\tau) \Psi(\tau)\,\tau^2  - \left( t \Psi(t) + \int_0^t  d\tau \,\Upsilon(t-\tau) \Psi(\tau)\, \tau \right)^2   \right]
\eea

\begin{figure}[t]
\includegraphics[width = 0.4\linewidth]{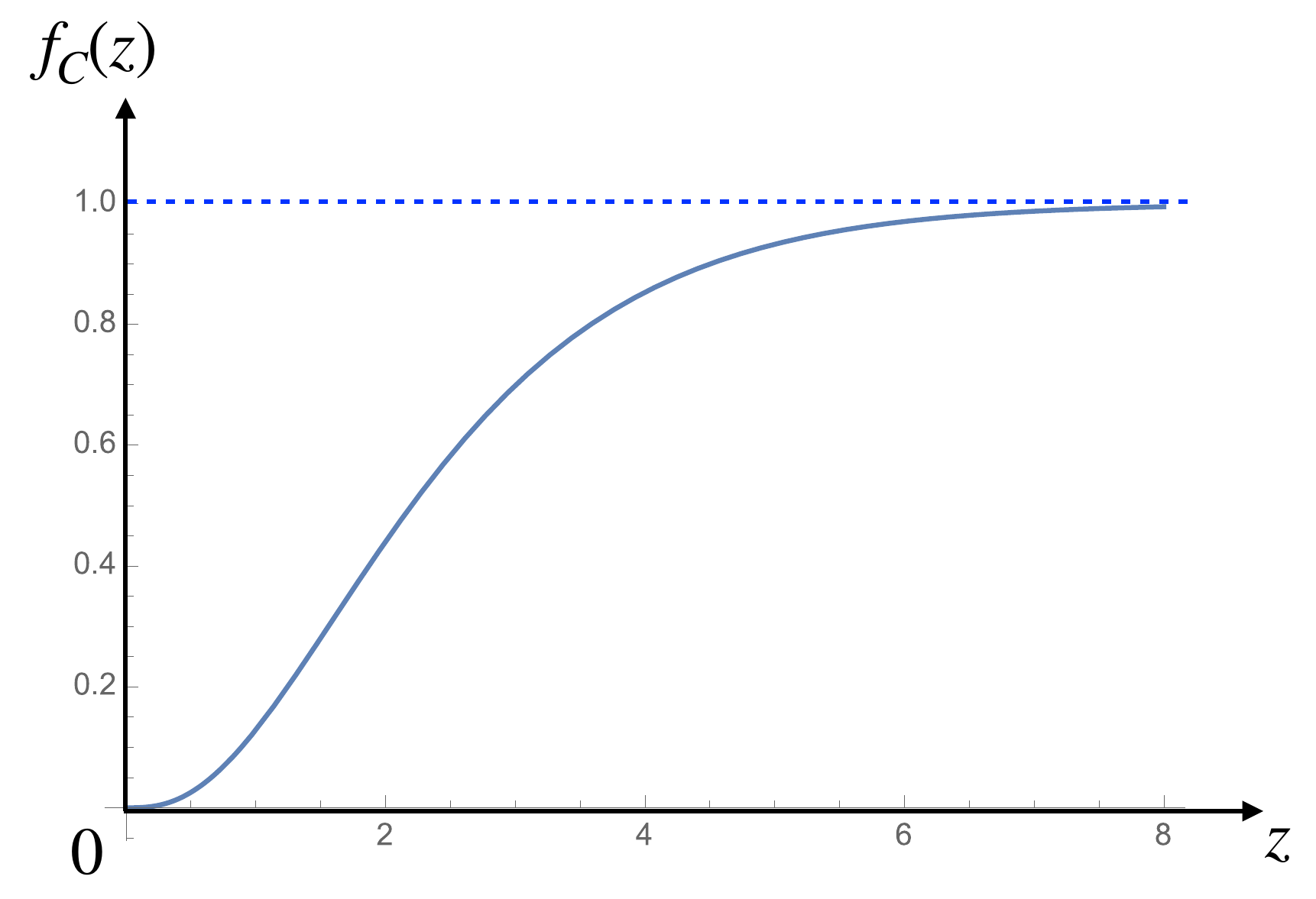}
\caption{Plot of the function $f_C(z)$ vs $z$, as given in Eq. (\ref{C2_exp}). The horizontal dashed line indicates the asymptotic value $f_C(z \to \infty) = 1$ [see Eq. (\ref{fC})].}\label{fig_fC}
\end{figure}
This result is valid for arbitrary interval distribution $\psi(\tau)$. However, it is useful to first consider the Poissonian resetting case, with $\psi(\tau) = r\,e^{-r\tau}$, where ${\cal C}_2(t)$ in Eq. (\ref{expr_C2}) can be computed explicitly. In this case, $\Psi(t) = e^{-r\,t}$. Furthermore, since $\tilde \psi(s) = r/(r+s)$, it follows from Eq. (\ref{def_upsilon}) that $\tilde \Upsilon(s) = r/s$, indicating $\Upsilon(t) =r$\, for all $t$. Evaluating simply the integrals in Eq. (\ref{expr_C2}) then gives
\bea \label{C2_exp}
{\cal C}_2(t) = \frac{4D^2}{r^2} f_{C}\left( r\, t\right) \quad, \quad f_C(z) = 1 - 2 z\,e^{-z} - e^{-2 z} \;.
\eea
The scaling function $f_C(z)$ is plotted in Fig. \ref{fig_fC} and has the asymptotic behaviors
\bea \label{fC}
f_C(z) \approx
\begin{cases}
&\frac{1}{3}\,z^3 \quad, \quad \hspace*{0.9cm}z \to 0 \\
& \\
&1 - 2 z\,e^{-z} \quad, \quad z \to \infty \;.
\end{cases}
\eea
Thus the correlation function $C_2(t)$ starts growing as $t^3$ at early times and finally saturates at long times to its stationary value $4D^2/r^2$. This stationary value was computed in Ref. \cite{gas_bosons} but not the temporal growth. For a general $\psi(\tau)$, while the expression for the correlator ${\cal C}_2(t)$ is explicit in Eq. (\ref{expr_C2}), it is not easy to evaluate it for all $t$. However, one can easily derive its asymptotic behaviors for small and large $t$ for a general $\psi(\tau)$, as we show below. 

\vspace{0.5cm}

\noindent{\it Early times.} From Eq. (\ref{expr_C2}), it is clear that, for small $t$, only the small $\tau$ behavior of $\psi(\tau)$ contributes. Let us assume that $\psi(\tau)$ has the following leading small $\tau$ behavior 
\bea \label{small_tau}
\psi(\tau) \approx A\, \tau^\nu \quad, \quad \nu > -1 \; \quad{\rm and} \; A > 0 \;.
\eea  
The Poissonian case corresponds to $\nu = 0$ and $A = r$. From Eq. (\ref{expr_C2}) one can show that the leading order behavior of ${\cal C}_2(t)$ for small $t$ is given by
\bea
{\cal C}_2(t) \approx \frac{4D^2\,A}{\nu +3}\, {t^{\nu + 3}} \quad, \quad t \to 0 \;.
\eea
For $\nu =0$ and $A=r$ (the Poissonian case), this result matches with Eqs. (\ref{C2_exp}) and the first line of (\ref{fC}). Thus the exponent $(\nu +3)$ for the power law growth of ${\cal C}_2(t)$ at early times depends only on the exponent $\nu$ that characterises the small $\tau$ behavior of $\psi(\tau)$. Indeed the ratio 
\bea \label{def_ratio}
\frac{{\cal C}_2(t)}{\psi(t)} \approx \frac{4 D^2}{\nu + 3}\, t^3 \;, \; t \to 0 \;,
\eea
grows universally as $t^3$ at early times, for any $\psi(\tau)$ with a short time behavior given in Eq. (\ref{small_tau}).

\vspace{0.5cm}

\noindent{\it Late times.} To analyse the late time behavior, we start from the Laplace transform of the JPDF given in Eq. (\ref{eq:Stat_state_laplace}). 
Multiplying Eq. (\ref{eq:Stat_state_laplace}) by $x_i^2 \, x_j^2$ with $i \neq j$ and integrating over all $x_i$'s gives
\bea \label{LT_C2}
\int_0^\infty dt\, e^{-st}\,\langle x^2_i(t) x_j^2(t)\rangle = \frac{4 D^2}{s\, \tilde \Psi(s)} \int_0^{\infty} dt\, e^{-st}\, \Psi(t) \, t^2 \;.
\eea
Similarly, one gets
\bea \label{LT_C2_2}
\int_0^\infty dt\, e^{-st}\,\langle x^2_i(t) \rangle = \frac{2 D}{s\, \tilde \Psi(s)} \int_0^{\infty} dt\, e^{-st}\, \Psi(t) \, t \;.
\eea 
To analyse the large time behavior of ${\cal C}_2(t)$, we need to inspect the small $s$ behavior of these expressions. This requires analysing $\tilde \Psi(s)$ for small $s$. The small $s$ behavior of $\tilde \psi(s)$ depends on the tail of $\psi(\tau)$ for large $\tau$. If $\psi(\tau)$ decays as 
\bea \label{def_gamma}
\psi(\tau)\sim \frac{B}{\tau^{1+\beta}} \quad, \quad \tau \to \infty 
\eea
with $\beta >0$. In this case, the small $s$ behavior of $\tilde \psi(s)$ is given by
\bea \label{small_s_psi}
\tilde \psi(s) \approx
\begin{cases}
&1 - \frac{B}{\beta}\, \Gamma(1-\beta)\, s^{\beta} \quad, \quad \beta < 1 \;, \\
& \\
& 1 - \langle \tau \rangle \, s \quad, \quad \quad \quad \hspace*{0.6cm}\beta > 1 \;,
\end{cases}
\eea 
with logarithmic corrections exactly at $\beta= 1$. Here we denote
\bea \label{av_tau}
\langle \tau \rangle = \int_0^\infty d\tau \tau \, \psi(\tau) = \int_0^\infty d\tau\, \Psi(\tau) \;,
\eea
which exists only for $\beta > 1$. Consequently from Eq. (\ref{res_Psi}), $\tilde \Psi(s)$ behaves as
\bea \label{small_s_Psi}
\tilde \Psi(s) \approx
\begin{cases}
& \frac{B}{\beta} \Gamma(1-\beta)\, s^{\beta-1} \quad, \quad 0<\beta < 1\\
& \\
& \langle \tau \rangle \quad, \quad \hspace*{2cm} \beta > 1
\end{cases}
\quad, \quad s \to 0 \;.
\eea
We substitute this small $s$ behavior in Eqs. (\ref{LT_C2}) and (\ref{LT_C2_2}) and analyse the small $s$ behavior of the Laplace transforms. It turns out that there are three cases which show distinct asymptotic behaviors to leading order for large $t$. Skipping details, one finds
\bea \label{sum_C2}
{\cal C}_2(t) \approx 
\begin{cases}
&2D^2 \beta(1-\beta)t^2 \quad, \quad 0<\beta<1 \\
& \\
& \frac{4D^2 B}{\beta \langle \tau\rangle(3-\beta)}\,t^{3-\beta}\quad, \quad 1<\beta<3 \\
& \\
&  {\cal C}_{2}(t \to \infty) \quad, \quad \quad \quad \quad \beta > 3 \;,
\end{cases}
 \quad, \quad t \to \infty \;.
\eea
where the constant ${\cal C}_{2}(t \to \infty)$ is given by
\bea \label{C2_inf}
{\cal C}_{2}(t\to \infty) = 4 D^2\, \left[\int_0^\infty d\tau\, h(\tau)\,\tau^2 - \left(\int_0^\infty d\tau\, h(\tau)\,\tau\right)^2 \right] \quad, \quad {\rm with} \quad h(\tau) = \frac{\Psi(\tau)}{\int_0^\infty \Psi(t)\,dt} \;.
\eea
Thus for $\beta < 3$, i.e., if the interval distribution $\psi(\tau)$ decays slower than $1/\tau^4$ for large $\tau$, the connected correlator ${\cal C}_2(t)$ keeps growing algebraically with increasing time $t$, while for $\beta>3$ it approaches a time independent stationary value at long times.

\section{Nonequilibrium stationary state and the relevant observables} \label{sec:NESS}

In this section, we first derive the joint distribution of the positions of the particles in the stationary state whenever it exists. We also define the main observables and outline the method for computing them in the stationary state, using a special solvable structure of the stationary joint distribution.

\subsection{Stationary joint distribution}

In the preceding section, we have studied the growth of the correlations between the positions of two particles as a function of time. We have seen that if $\psi(\tau)$ decays as a power law $\psi(\tau)\sim B/\tau^{\beta +1}$ for large $\tau$, then the connected correlator ${\cal C}_2(t)$ approaches a stationary form only for $\beta > 3$. It turns out, however, that the full joint distribution reaches a stationary form already for $\beta > 1$. For $1<\beta < 3$, while the full joint distribution still has a stationary form at late times, its second moment diverges, leading to a divergence of ${\cal C}_2(t)$ as $t \to \infty$. In this section, we compute the full time independent joint distribution, valid for all $\beta > 1$.   

To derive the stationary state ${\mathcal{P}^*}(\vec{x})$, when it exists, we start from the Laplace transform of the full JPDF in Eq.~(\ref{eq:Stat_state_laplace}). For the stationary state to exist, the Laplace transform of the JPDF must behave for small $s$ as
\bea  \label{small_s}
\tilde{\mathcal{P}}(\vec{x},s|\vec{x}_0)  = \int_0^\infty dt\, e^{-st} {\mathcal{P}}(\vec{x},t|\vec{x}_0) \xrightarrow{s \to 0}  \frac{{\mathcal{P}^*}(\vec{x})}{s} \;.
\eea
The RHS of Eq. (\ref{eq:Stat_state_laplace}) behaves, for small $s$, as
\bea \label{small_s_rhs}
{\rm RHS}  \xrightarrow{s \to 0} \frac{1}{s}\, \frac{\int_0^{\infty} d\tau\, \Psi(\tau)\, \prod_{i=1}^{N}\frac{1}{\sqrt{4\pi D \tau}}\exp\left(-\frac{x_i^2}{4D\tau}\right)}{\int_0^{\infty} d\tau\, \Psi(\tau)} \;,
\eea
provided the integral $\int_0^\infty d\tau\, \Psi(\tau)$ is finite. For example, for the distribution with a power law tail $\psi(\tau) \sim B/\tau^{\beta+1}$ for large $\tau$, one must have $\beta > 1$ for the integral to converge. Thus, comparing both the left and right hand side of Eq. (\ref{eq:Stat_state_laplace}) as $s \to 0$, we find that the stationary state, whenever $\int_0^\infty d\tau\, \Psi(\tau)$ is finite, exists and is given by
\begin{equation}\label{eq:stat_state}
 \mathcal{P}^*(\vec{x})  =\frac{\int_0^{\infty} d\tau\, \Psi(\tau)\, \mathcal{G}_0(\vec{x},\tau)}{\int_0^{\infty} d\tau\, \Psi(\tau)}\,,
\end{equation} 
where $\mathcal{G}_0(\vec{x},\tau)$ is given in Eq. (\ref{eq:propagator_factorizable}). The stationary JPDF can be expressed in a convenient way by 
defining a normalised distribution function 
\begin{equation}\label{eq:h(t)}
    h(\tau) = \frac{\Psi(\tau)}{\int_0^{\infty} d\tau\, \Psi(\tau)}\,,
\end{equation}
in terms of which the JPDF in Eq. (\ref{eq:stat_state}) reads
\begin{equation}\label{eq:SS_ciid_conh}
\mathcal{P}^*(\vec{x}) = \int_0^{\infty} d\tau\, h(\tau)\, \mathcal{G}_0(\vec{x},\tau) \;, 
\end{equation}
where we recall once again that $\mathcal{G}_0(\vec{x},\tau) = \prod_{i=1}^{N}\frac{1}{\sqrt{4\pi D \tau}}e^{-\frac{x_i^2}{4D\tau}}$. The first remark that we want to make is that the stationary JPDF $\mathcal{P}^*(\vec{x})$ does not factorise into a product of marginal distributions of $x_i$'s. This indicates that these variables are correlated in the stationary state. This was also evident from the calculation of the connected correlator ${\cal C}_2(t)$ in the previous section (whenever it exists). These correlations are attractive and all-to-all and get dynamically generated due to the {\it simultaneous resetting} of all the particles. This stationary JPDF in Eq. (\ref{eq:SS_ciid_conh}) has a CIID structure which has been recently found in a number of models, both classical and quantum~\cite{sabhapandit2024noninteracting,kulkarni2025dynamically,dean2025exact,mesquita2025dynamically,biroli2025exp,mesquita2025dynamicallyb}. This means the following: one can think of $\tau$ as a random variable drawn from the PDF $h(\tau)$ (since $\int_0^\infty d\tau\, h(\tau) = 1$ and $h(\tau) > 0$ for all $\tau>0$, one can interpret $h(\tau)$ as the PDF of $\tau$). Once $\tau$ is fixed, then the 
$x_i$'s are IID variables distributed via $\mathcal{G}_0(\vec{x},\tau)$ which has a factorized form and finally one has to integrate over $\tau$, drawn from $h(\tau)$. 
We will see below that, although the positions are strongly correlated in the stationary state, this CIID structure allows to compute several physical observables in terms of the single function $h(\tau)$ defined in Eq. (\ref{eq:h(t)}). Note that for the Poissonian case 
\bea \label{exp_case}
\Psi(\tau) = e^{-r \tau} \quad, \quad {\rm and} \quad h(\tau) = r\,e^{-r\tau} \;,
\eea
and hence the JPDF in Eq. (\ref{eq:SS_ciid_conh}) reduces to Eq. (\ref{eq:SS_poissonian}). 

Let us also remark that the stationary state characterized by the JPDF $\mathcal{P}^*(\vec{x})$ is a nonequilibrium steady state (NESS) and is entirely determined by the dynamics itself, rather than by an energy function as in equilibrium systems.  
We also note that this derivation does not use the explicit form of $\mathcal{G}_0(\vec{x},t)$ and in fact this formula for the stationary JPDF is very general and holds for any process, not necessarily diffusive. The propagator $\mathcal{G}_0(\vec{x},t)$ will differ from one process to another but the stationary state is uniquely determined by this propagator and the distribution $\psi(\tau)$ of the intervals between resettings. Thus one can generate a whole class of stationary states by varying either the distribution $\psi(\tau)$ or the propagator of the process itself. The only important assumption is that the particles move independently between two resettings. Thus one can {\it stochastically control} or tune, using either $\psi(\tau)$ or $\mathcal{G}_0(\vec{x},t)$, the form of the stationary JPDF and also the correlations between the particles in the stationary state, whenever it exists.

\subsection{Relevant observables}

Knowing explicitly the JPDF in Eq. \eqref{eq:SS_ciid_conh} in terms of $h(\tau)$, several observables can be computed, first for free Brownian particles for fixed $\tau$ and then averaging over $\tau$ drawn from the PDF $h(\tau)$. Below we list few of such observables.

\vspace*{0.5cm}
\noindent{\it $\bullet$ Average density and correlations.} The average density, normalized to unity, is defined as
\begin{eqnarray}
    \rho_N(x) = \frac{1}{N} \left\langle \sum_{i=1}^{N} \delta(x - x_i) \right\rangle\,,
\end{eqnarray}
which measures the average fraction of particles inside the interval $[x,x+dx]$. Due to the symmetry between particles, $\rho_N(x)$ coincides with the one-point marginal of the stationary distribution, namely
\begin{equation}\label{eq:rhodef}
    \rho_N(x)=\mathcal{P}_{\mathrm{stat}}(x) = \int_{-\infty}^{\infty} dx_1 \cdots \int_{-\infty}^{\infty} dx_{N-1} \, \mathcal{P}^*(\vec{x})=\int_0^{\infty} d\tau\, h(\tau)\, p_0(x \vert \tau) \,,
\end{equation}
where $p_0(x \vert \tau)$ is the single particle propagator from $0$ to $x$ in time $\tau$. For the Brownian particles that we focus on in this paper, we simply have
\bea \label{p0_BM}
p_0(x \vert \tau) = \frac{1}{\sqrt{4 \pi D \tau}}\, e^{-\frac{x^2}{4 D\,\tau}} \;.
\eea
Interestingly the global average density in Eq. (\ref{eq:rhodef}) is completely independent of $N$. Hence it does not carry any information about the correlations between the particles. To detect the correlations in the stationary state, one needs to compute the two-point correlator ${\cal C}_2(t \to \infty)$ which was already computed in terms of $h(\tau)$ in Eq. (\ref{C2_inf}). 

\vspace*{0.5cm}
\noindent{\it $\bullet$ Order statistics.} To probe the local statistics of this gas of strongly correlated particles on the line, it is convenient to order the particle positions $x_i$'s in decreasing order and relabel them as $\{M_1 > M_2 > \dots > M_N\}$, where $M_1 = \max\{x_1, x_2, \cdots, x_N\}$ and $M_N= \min\{x_1, x_2, \cdots, x_N\}$. This allows us to study quantities such as the probability $\text{Prob}\{M_k = \omega\}$ that the $k^{\rm th}$ rightmost particle is at position $\omega$.  Thanks to the CIID structure, as noted above, we can write
\begin{equation} \label{PDF_Mk}
    \text{Prob}[M_k = \omega]=\int_0^{\infty}d\tau\,h(\tau)\text{Prob}[M_k(\tau) = \omega] \;,
\end{equation}
where $M_k(\tau)$ is the maximum of $N$ independent Brownian motions at time $\tau$, or equivalently that of $N$ independent Gaussian 
variables with zero mean and variance $2D\tau$. The statistics of $M_{k}$ and $M_{N-k}$ are related (in law) by the symmetry
\bea \label{sym}
M_{k} \equiv - M_{N-k } \;.
\eea
This follows by changing $x \to -x$ and using the symmetry of $\mathcal{P}^*(\vec{x})$ under this inversion.
Since we are interested in the large $N$ limit, we set $k = \alpha\,N$ where $0 \leq \alpha \leq 1$. For $\alpha = O(1)$, one then probes the bulk of the gas near the origin, while for $\alpha = O(1/N)$, i.e., $k=O(1)$, we probe instead the gas at its right edge.
Without loss of generality we consider $0< \alpha < 1/2$ (the case $1/2<\alpha<1$ follows by the inversion symmetry (\ref{sym})). The case $\alpha = 1/2$ is critical and follows a slightly different treatment \cite{mesquita2025dynamically}. When $0<\alpha<1/2$, the large $N$ behavior of $\text{Prob}[M_k(\tau) = \omega]$ in Eq.~(\ref{PDF_Mk}) for $N$ IID random variables, each drawn from $p_0(x \vert \tau)$ in Eq. (\ref{p0_BM}), was analysed in great detail in Ref. \cite{gas_bosons}. It was found that, to leading order for large $N$, the probability $\text{Prob}[M_k(\tau) = \omega]$ can be approximated simply by a delta-function 
\bea \label{kmax_IID}
\text{Prob}[M_k(\tau) = \omega] \underset{N \to \infty}{\longrightarrow} \delta(\omega - \omega^*(\alpha,\tau)) \;,
\eea 
where $\omega^*(\alpha, \tau)$ is defined as the $\alpha$-quantile, i.e., 
\begin{equation} \label{def_alpha}
\int_{\omega^*(\alpha, \tau)}^{+\infty} p_0(x \vert \tau) \, \mathrm{d}x = \alpha \;.
\end{equation}
Thus $\omega^*(\alpha, \tau)$ is the value of $x$ such that there are, on average, $\alpha N$ variables above $\omega^*(\alpha, \tau)$. In the Brownian case \blue{$p_0(x \mid \tau)=\frac{1}{\sqrt{4\pi D \tau}} \exp\!\left(-\frac{x^2}{4D\tau}\right) $} and the quantile then is explicitly given by
\bea \label{quant_BM}
\omega^*(\alpha, \tau)= \sqrt{4D\tau}\: \text{erfc}^{-1}(2\alpha) \quad, \quad {\rm with} \quad 0 < \alpha < 1/2 \;.
\eea
\blue{where $\operatorname{erfc}(z) = \frac{2}{\sqrt{\pi}} \int_{z}^{\infty}du\, e^{-u^{2}}\, $ and $\operatorname{erfc}^{-1}(z)$ is its inverse function.}
Substituting this result in Eq.~(\ref{quant_BM}) in (\ref{PDF_Mk}) leads to the leading large $N$ behavior (for fixed $0<\alpha < 1/2$)
\begin{equation}\label{eq:max_delta}
    \text{Prob}[M_{k} = w]\underset{N \to \infty}{\longrightarrow} \int \mathrm{d}\tau \, h(\tau) \, \delta \left( w - \omega^*(\alpha, \tau) \right) \;.
\end{equation}
Note that $\omega^*(\alpha, \tau) >0$ for any $0 < \alpha < 1/2$ and hence, from Eq. (\ref{eq:max_delta}), we see that the limiting distribution of $M_k$ is supported over the positive semi-axis only. This result in Eq. (\ref{eq:max_delta}) generalises the result for the Poissonian protocol $h(\tau) = r\, e^{-r \tau}$ obtained in \cite{Exact_Extreme}.

\vspace*{0.5cm}
\noindent{\it $\bullet$ Gap statistics.} Another interesting microscopic observable in a correlated gas is the spacing between two successive particles. This is denoted by the gap $d_k = M_{k} - M_{k+1}$ where $M_k$ denotes the position of the $k^{\rm th}$ particle. Let $\text{Prob}[d_k = g]$ denote the distribution of the $k^{\rm th}$ gap. It can be expressed as  
\begin{equation} \label{eq:d_k_iniziale}
\text{Prob}[d_k = g] =  \int_{0}^{\infty} d\tau \, h(\tau) \, \text{Prob}[d_k(\tau) = g] \;,
\end{equation}
where $d_k(\tau)=M_k(\tau)-M_{k+1}(\tau)$ is the distance between the $k^{\rm th}$ and $(k+1)^{\rm th}$ maxima of a set of $N$ IID Gaussian random variables of zero mean and variance $2 D \tau$. For fixed $\tau$, we just need to compute $\text{Prob}[d_k(\tau) = g]$ for $N$ IID variables, each drawn from $p_0(x,\vert \tau)$ in Eq. (\ref{p0_BM}). This was already done in great detail in the limit of large $N$ using a saddle point method in Refs.~\cite{gas_bosons,Exact_Extreme}. It was found that 
\bea \label{gap_IID}
\text{Prob}[d_k(\tau) = g]  \underset{N \to \infty}{\longrightarrow}  N p_0(\omega^*\vert \tau) e^{-N\,p_0(\omega^*\vert \tau)g} \;,
\eea
where $k=\alpha N$, $\omega^*\equiv \omega^*(\alpha,\tau)=\sqrt{4D\, \tau}\,\text{erfc}^{-1}(2\alpha)$ and $p_0(\omega^* \vert \tau)=\frac{1}{\sqrt{4\pi D \tau}}\exp\left( -[\text{erfc}^{-1}(2\alpha)]^2\right)$. Substituting this result in Eq. (\ref{eq:d_k_iniziale}) we then get
\begin{equation}\label{eq:d_k_p(omega)}
    \text{Prob}[d_k = g] \approx N\, \int_{0}^{\infty} d\tau \, h(\tau) \,p_0(\omega^*\vert\tau) e^{-N\,p_0(\omega^*\vert\tau)g} \;.
\end{equation}
This generalises the result for the Poissonian resetting protocol $h(\tau) = r\,e^{-r \tau}$ in Ref. \cite{Exact_Extreme}. 

\vspace*{0.5cm}
\noindent{\it $\bullet$ Full counting statistics and variance.}
We also study the full counting statistics (FCS), i.e., the probability $P(N_L, N)$ of finding $N_L$ particles out of $N$ within a symmetric interval $[-L, L]$ centered around the resetting point $x = 0$. To compute it for our gas, we first fix $\tau$ and define the random variable $N_L(\tau)$ as the number of particles in the interval $[-L,+L]$. Since, for fixed $\tau$, the particles are independent, the distribution of $N_L(\tau)$ follows a simple binomial law
\bea \label{NL_IID}
{\rm Prob.}(N_L(\tau) = n) = \binom{N}{n} [q(\tau)]^{n} [1 - q(\tau)]^{N-n} \;,
\eea
where $q(\tau) = \int_{-L}^{L} p(y,\tau) \, dy = \operatorname{erf} \left( \frac{L}{\sqrt{4D\tau}} \right)$ is the probability for a Brownian particle 
to be inside the interval $[-L,L]$ at time $\tau$. Finally, averaging over $\tau$, drawn from $h(\tau)$, the distribution $P(N_L,N)$ is given by
\begin{equation}\label{eq:fulcstat}
    P(N_L,N) = \int_{0}^{\infty} d\tau \, h(\tau) \binom{N}{N_L} [q(\tau)]^{N_L} [1 - q(\tau)]^{N-N_L} \;.
\end{equation}
As $N$ increases, the binomial distribution inside the integral \eqref{eq:fulcstat} becomes sharply peaked around its average $N\,q(\tau)$. It turns out that, to leading order for large $N$, it is sufficient to keep only the mean value and ignore the fluctuations around the
mean~\cite{gas_bosons}, i.e., 
\bea \label{FCS_IID}
\binom{N}{N_L} [q(\tau)]^{N_L} [1 - q(\tau)]^{N-N_L}  \underset{N \to \infty}{\longrightarrow}  \delta(\kappa - q(\tau)) \;,
\eea
where $\kappa = N_L/N$. Substituting this result in Eq. (\ref{eq:fulcstat}) one gets
\begin{equation}\label{eq:fullcount_tot}
    P(N_L,N) \approx \frac{1}{N} H\left( \kappa = \frac{N_L}{N}\right) {\quad}, \quad {\rm where} \quad H(\kappa)= \int_{0}^{\infty} d\tau \, h(\tau) \delta(\kappa - q(\tau)) \;.
\end{equation}
This generalises the result for the Poissonian resetting protocol $h(\tau) = r\,e^{-r \tau}$ in Ref. \cite{Exact_Extreme}. 

In many applications, the number variance, i.e., the second cumulant of the FCS, namely
\bea \label{def_Var}
{\rm Var}(N_L) = \langle N_L^2 \rangle - \langle N_L \rangle^2 \;, 
\eea
is the most important characterization of the fluctuations that can be easily measured in experiments \cite{torquato2018hyperuniform}. In particular, it is interesting to investigate how the variance of $N_L$ depends on the size $L$ of the interval. For fixed $\tau$, the distribution of $N_L(\tau)$ for $N$ IID random variables, each drawn from $p_0(x \vert \tau)$, is a simple binomial distribution as in Eq. (\ref{NL_IID}). Consequently, for fixed $\tau$, the variance is simply $N q(\tau) [1 - q(\tau)$. Hence averaging over $\tau$ we find
\bea \label{def_Var2}
{\rm Var}(N_L) =  N\, \int_0^\infty d\tau \, h(\tau) \, q(\tau) [1 - q(\tau)] = N\, \int_0^\infty d\tau \, h(\tau)\, {\rm erf}\left(\frac{L}{\sqrt{4 D \tau}}\right)\, {\rm erfc}\left(\frac{L}{\sqrt{4 D \tau}}\right) \;,
\eea
where we used $q(\tau) = {\rm erf}(L/\sqrt{4 D \tau})$.

\vspace*{0.5cm}

In the two following sections, we study these observables in detail for different resetting protocols $h(\tau)$. One can then generate a wide variety of rich behaviors for these observables, depending on $h(\tau)$. Some features of these distributions are universal, i.e., independent of the detail of $h(\tau)$. For example, the behaviors for small values of the arguments of $\rho_N(x)$, $\text{Prob}[M_k = \omega]$, and $\text{Prob}[d_k = g]$ are independent of the form of $h(\tau)$. A similar universal behavior appears in the FCS when taking the limit \( N_L \to N \). Conversely, in the large-$x$, large-$\omega$, large-$g$ or small-$N_L$ regime, the observables depend only on the large-$\tau$ tail of $h(\tau)$, and we have derived explicit results for three general classes of tail behaviors.
Since the observables $M_k$ and $d_k$ have been widely studied in the literature for IID variables \cite{AFirstCourseinOrderStatistics,OrderStatistics,StatisticsofExtremesandRecordsinRandomSequence,Extremevaluestatisticsofcorrelatedrandomvariables:Apedagogicalreview}, we compare some of these classical results to our case in order to see how correlations change these observables in this case.
We illustrate three concrete cases for $\psi(\tau)$ in Section \ref{sec:three_examples} before giving the general results in Section \ref{sec:risultati}.

\section{Analytical results for three particular resetting protocols}\label{sec:three_examples}

We now study the behavior of the system for three specific examples of the resetting time distribution $\psi(\tau)$. This will provide concrete insight before generalizing the results to an arbitrary $h(\tau)$ in Section~\ref{sec:risultati}. We begin by recalling the well-studied Poissonian case, for which $h(\tau)=r\,e^{-r\tau}$, and then examine two qualitatively different protocols $h(\tau)$: one with a power-law tail and one with a bounded support. These two cases were chosen because they exhibit significantly different behaviors even for IID behaviors (i.e., for fixed $\tau$), leading to distinct outcomes for all the observables of interest.

\subsection{Poissonian resetting protocol}

We begin with the Poissonian resetting case introduced in Section \ref{sec:introduction}, corresponding to the inter-reset time distribution $\psi(\tau) = r e^{-r \tau}$. Physically, this distribution arises because, in each small interval $dt$, particles reset with probability $r\,dt$ and diffuse otherwise.
This case was studied in detail in Refs. \cite{gas_bosons,Exact_Extreme}. Here, we recall these derivations for the sake of completeness and for comparison to other protocols. In this case, $h(\tau)$ coincides with the inter-reset distribution $\psi(\tau) = r\,e^{-r \tau}$, i.e., 
\blue{\bea \label{h_Poisson}
h(\tau) = \psi(\tau) =   r\,e^{-r \tau}\;.
\eea}
Using $\Psi(\tau) = e^{-r \tau}$, the stationary JPDF in Eq. (\ref{eq:stat_state}) reads 
\begin{equation}\label{eq:stat_state_exp}
 \mathcal{P}^*(\vec{x})  =r\, \int_0^{\infty} d\tau\, e^{-r\tau} \prod_{i=1}^{N}\frac{1}{\sqrt{4\pi D \tau}} \exp\left(-\frac{x_i^2}{4D\tau}\right)\,.
\end{equation} 
We now consider the different observables introduced earlier.

\vspace*{0.5cm}
\noindent{\it $\bullet$ Average density and correlations.}
The average density reads, from Eq. (\ref{eq:rhodef}), 
\begin{equation} \label{rho_Poisson}
    \rho_N(x) = r \int_0^\infty d\tau \, e^{-r \tau} \frac{1}{\sqrt{4 \pi D \tau}}\, e^{-\frac{x^2}{4 D \tau}} = \frac{1}{l} R\left( \frac{x}{l}\right)
\end{equation}    
where \(l=\sqrt{\frac{D}{r}}\) denotes the typical distance that a single particle travels between two consecutive resettings and the scaling function $R(z)$ is given by
\bea \label{Rofz}
R(z) = \frac{1}{2}\, e^{-|z|} \;.
\eea     
This scaling function coincides with that of the single particle position distribution with Poissonian resetting, derived originally in Ref. \cite{DiffusionwithStochasticResetting}. Thus, the density decays exponentially with the characteristic length scale \(l=\sqrt{\frac{D}{r}}\). This prediction has been experimentally verified using colloidal particles as the diffusing agents and optical tweezers to implement the resetting mechanism \cite{Experimental_1}.

Although the particles diffuse independently between resets, the resetting mechanism introduces strong correlations. This is evident from the fact that the JPDF does not factorize. Indeed, in this case the full time dependent correlator ${\cal C}_2(t)$ can be computed explicitly for all $t$ (see Eq. (\ref{C2_exp})). In particular, in the stationary state, i.e., when $t \to \infty$, the correlator is given by 
\bea \label{eq:correlator_poissonian}
{\cal C}_2(t \to \infty) = \frac{4D^2}{r^2} \;,
\eea
which indicates strong correlations at all distances. We recall that these correlations arise purely from the dynamics of the system, as the particles do not interact directly.

\vspace*{0.5cm}
\noindent{\it $\bullet$ Order statistics.}  For fixed $\tau$, the order and extreme value statistics for $N$ IID random variables, each drawn from $p_0(x \vert \tau)$ in Eq. (\ref{p0_BM}), was discussed briefly in the previous section. We discuss it here in more detail for different ranges of $k$, that include both the edge, where $k = O(1)$ and the bulk, where $k = O(N)$. From the standard theory of extreme value statistics of IID variables, 
it is known that, for large $N$, this random variable $M_k(\tau)$ can be expressed as \cite{gas_bosons} 
\bea \label{kthIID}
M_k(\tau) \approx \sqrt{4\,D\,\tau\,\ln N} + \sqrt{\frac{D\,\tau}{\ln N}}\, \chi_k \;,
\eea
where $\chi_k$ has a generalized Gumbel distribution, independent of $N$, and is given by
\bea \label{kGumbel}
{\rm Prob}(\chi_k = y) = \frac{1}{(k-1)!}\, e^{-k\,y-e^{-y}} \;.
\eea 
The first term in Eq. (\ref{kthIID}) dominates for large $N$ and is deterministic. The fluctuations are encoded in the second term, which, however, is subdominant for large $N$. Hence, Eq. (\ref{PDF_Mk}) can be approximated at leading order for large $N$ by (\ref{eq:max_delta}). For $h(\tau)=r\,e^{-r\tau}$, the integral in Eq.~\eqref{eq:max_delta} can be performed explicitly \cite{gas_bosons} and it gives, for $0<\alpha<1/2$
\begin{equation} \label{Mk_edge}
    \text{Prob}[M_{k} = \omega] \approx \frac{1}{L_N}\, S \!\left(\frac{\omega}{L_N} \right)\,,
\end{equation}
where $L_N = \sqrt{\frac{4D\ln N}{r}}$ for any $k=O(1)$ and the scaling function is given by
\begin{equation}\label{eq:S(z)}
    S(z)=2ze^{-z^2}\,\theta(z) \;,
\end{equation}
with $\theta(z)$ denoting the Heaviside step function. 

In contrast, when we probe the bulk by setting $k = \alpha \, N$ with $\alpha = O(1)$, we consider the range $0<\alpha<1/2$. In this case, one can express the random variable $M_k(\tau)$ for fixed $\tau$ as \cite{gas_bosons} 
\bea \label{Mk_Gauss}
M_{k = \alpha N}(\tau) \approx \sqrt{4 D\,\tau}\, {\rm erfc}^{-1}(2\alpha) + \sqrt{\frac{\alpha(1-\alpha)}{N}} \frac{1}{ p_0(\omega^* \vert \tau)}\, N(0,1)
\eea
where $p_0(x \vert \tau) = e^{-x^2/(4 D\,\tau)}/\sqrt{4 \pi D \, \tau}$ and we recall that $\omega^*\equiv \omega^*(\alpha,\tau)=\sqrt{4D\, \tau}\,\text{erfc}^{-1}(2\alpha)$. The quantity $N(0,1)$ denotes a Gaussian random variable of zero mean and unit variance. As long as $0<\alpha<1/2$, 
we have ${\rm erfc}^{-1}(2\alpha)>0$ and, hence, the random variable $M_{k = \alpha N}(\tau)$ in Eq. (\ref{Mk_Gauss}) is dominated by the first term and one can ignore the second term representing the fluctuations for large $N$. In this case, averaging over $h(\tau)$ in Eq. (\ref{PDF_Mk}), one finds that, for large $N$, the distribution has the scaling form  
\begin{equation} \label{Mk_bulk}
    \text{Prob}[M_{k = \alpha N} = \omega] \approx \frac{1}{\Lambda(\alpha)}\, S \!\left(\frac{\omega}{\Lambda(\alpha)} \right)\,,
\end{equation}
where the scale factor $\Lambda(\alpha)$, for $0<\alpha<1/2$, is given by 
\begin{equation}\label{eq:lambd_alpha_def}
    \Lambda(\alpha)=\sqrt{\frac{4D}{r}}\,\text{erfc}^{-1}(2\alpha)\quad \text{with}\quad \alpha=\frac{k}{N}\,.
\end{equation}
The scaling function $S(z)$ is exactly the same as in Eq. (\ref{eq:S(z)}). Thus, while the scale factor changes from the edge to the bulk, the scaling function remains the same for all $0<\alpha<1/2$. For $1/2<\alpha<1$, the distribution is supported over $z \in (-\infty,0]$ with the scaling function $S(z) = 2 |z|\,e^{-z^2}\,\theta(-z)$.

Exactly at $\alpha = 1/2$, i.e., for the middle particle, one finds ${\rm erfc}^{-1}(2 \alpha) = {\rm erfc}^{-1}(1)=0$. Hence,  
from Eq. (\ref{Mk_Gauss}), we see that the leading term vanishes identically for $\alpha = 1/2$ and, consequently, $M_{k = N/2}(\tau)$ is dominated
by the second term, representing the fluctuations. Hence, the distribution of $M_{k=N/2}$, averaged over $h(\tau)$ in Eq. (\ref{PDF_Mk}) yields a completely different scaling behavior~\cite{mesquita2025dynamically}
\bea \label{middle}
\text{Prob}[M_{k = N/2} = \omega] \approx \sqrt{\frac{r\,N}{2 D}}\, R \left( \sqrt{\frac{r\, N}{2 D}}\, \omega \right) \;,
\eea
where the scaling function $R(z)$ is exactly as in Eq. (\ref{Rofz}). Thus the position of the middle particle typically scales as $1/\sqrt{N}$, which means that it has much suppressed fluctuations compared to other particles. Moreover, the scaling function $R(z)$ coincides with the scaling function describing the global average density in Eq.~(\ref{rho_Poisson}). This is actually a general fact for many models with a CIID structure and with Poissonian resetting~\cite{mesquita2025dynamically}.

\vspace*{0.5cm}
\noindent{\it $\bullet$ Gap statistics.}
 The gap distribution in the large $N$ limit can also be explicitly computed and for $0<\alpha<1/2$ one finds~\cite{gas_bosons}
 \begin{equation}
    \text{Prob}[d_k = g] \approx \frac{1}{\lambda_N(\alpha)}\, D \!\left(\frac{g}{\lambda_N(\alpha)} \right)\,,
\end{equation}
where 
\begin{equation}
    D(z)=2\int_0^\infty du\,\exp\!\left(-u^2-\frac{z}{u}\right)\,,
\end{equation}
and the scale factor is 
\begin{equation}\label{eq:lambdaN_definition}
    \lambda_N(\alpha)=\frac{1}{Nb\sqrt{r}}\quad \text{with}\quad b = \frac{\exp\left(-\left[\text{erfc}^{-1}(2\alpha)\right]^2\right)}{\sqrt{4\pi D}}\,.
\end{equation}
This result holds both in the bulk and near the edges. Note that \(\lambda_N(\alpha)\) is \(O(1/N)\) in the bulk, while at the edges it behaves as \(\lambda_N(\alpha)\to l_N=\sqrt{\frac{D}{rk^2 \ln N}}\), indicating an inhomogeneous gas: particles are very closely packed in the bulk and much sparser at the boundaries.
The scaling function $D(z)$ behaves as
\begin{equation} \label{D_asympt_Pois}
    D(z) \approx 
\begin{cases}
\sqrt{\pi} + 2z \ln z \;, & \quad z \to 0 \\
& \\
2 \sqrt{\dfrac{\pi}{3}} \exp\left[-3 \left(\dfrac{z}{2}\right)^{2/3} \right]  \;, & \quad z \to \infty \;.
\end{cases}
\end{equation}
Thus the gap distribution $D(z)$ decays as a stretched exponential $\sim e^{-c z^{2/3}}$ for large $z$, which is slower than exponential.
Had they been IID variables, then the associated gap distribution would have an exponential tail. This slower than exponential tail, in the present case, is a fingerprint of strong correlations between the positions of the particles. In addition, $D(z)$ approaches a nonzero constant $\sqrt{\pi}$ as $z \to 0$. This is different from the well known Wigner-Dyson spacing distribution for the eigenvalues of Gaussian random matrix ensembles, where the gap distribution vanishes as $z \to 0$, indicating eigenvalue repulsion~\cite{mehta2004random, forrester2010log}. In our case, there is no repulsion between the particles since $D(0) > 0$. Instead, the fact that $D(0)>0$ indicates an all-to-all attraction between the particles. 

\begin{figure}[t]
\includegraphics[width = 0.4\linewidth]{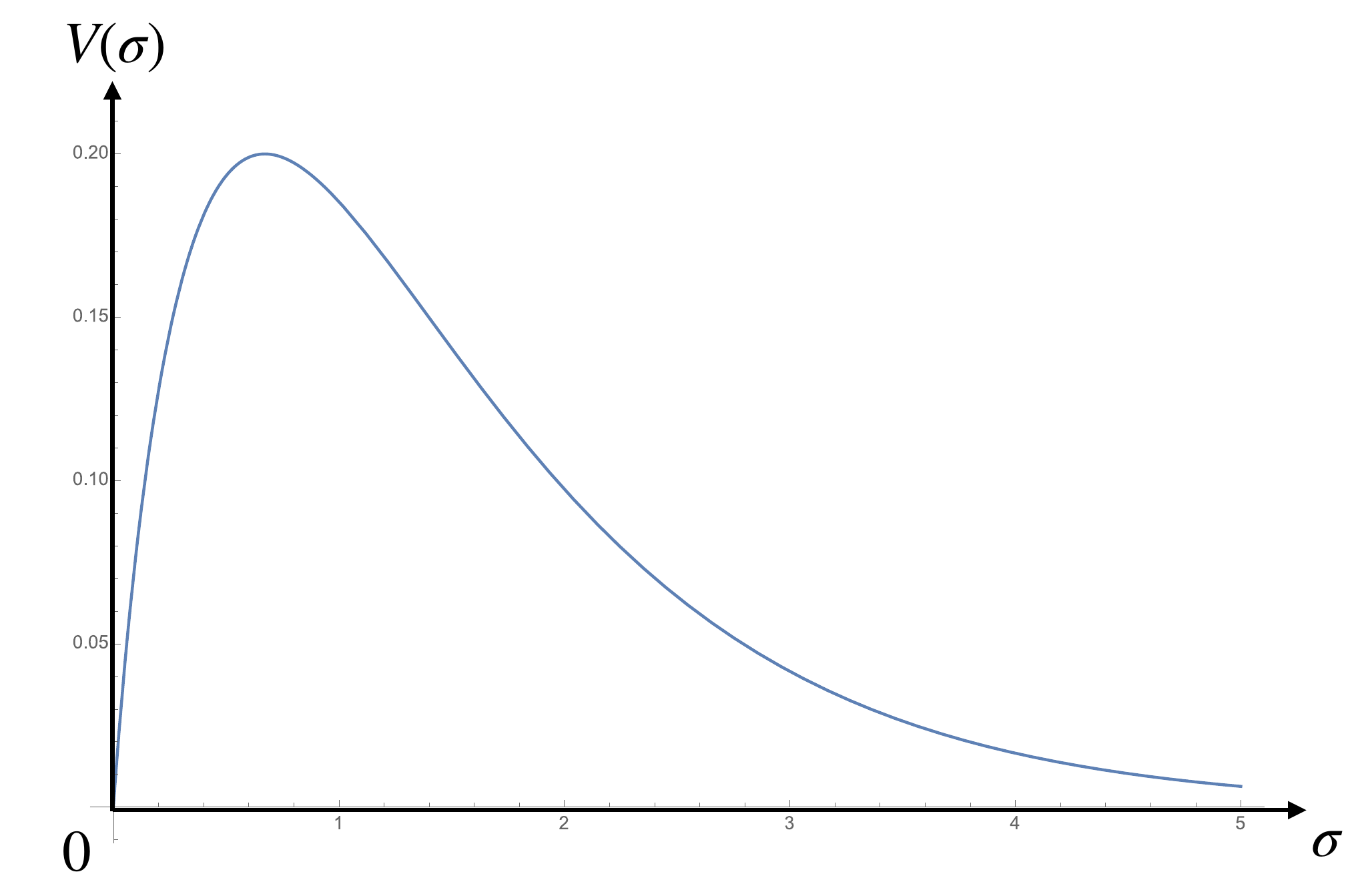}
\caption{Plot of the function $V(\sigma)$ vs $\sigma$, as given in Eq. (\ref{V_gamma0}).}\label{fig_V}
\end{figure}
\vspace*{0.5cm}
\noindent{\it $\bullet$ Full counting statistics and variance.} 
Finally, the full counting statistics (FCS) can also be derived by exploiting the CIID structure from Eqs. (\ref{eq:fulcstat})-(\ref{eq:fullcount_tot}) and the scaling function $H(\kappa)$ is given by \cite{gas_bosons}
\begin{equation}\label{eq:H_poissonian}
   H(\kappa) = \sigma \sqrt{\pi} \, [u(\kappa)]^{-3} \exp\!\left( -\frac{\sigma}{[u(\kappa)]^2} + [u(\kappa)]^2 \right)\,, 
\end{equation}
where $\sigma = rL^2/(4D)$ and \(k=\frac{N_L}{N}\) represents the fraction of particles inside the box \([-L,L]\).
\blue{The function $u(\kappa) = \text{erf}^{-1}(\kappa)$ is the inverse function of the error function defined by $\operatorname{erf}(u) = \frac{2}{\sqrt{\pi}} \int_0^u e^{-t^2} \, dt$} . The asymptotics of the above equation are 
\blue{\begin{equation}\label{eq:asympt_H_poiss}
  H(\kappa) \sim 
\begin{cases}
\displaystyle \frac{8\sigma}{\pi \kappa^3} \exp\!\left(-\frac{4\sigma}{\pi \kappa^2}\right), & \kappa \to 0 \\[10pt]
\displaystyle \frac{\sigma }{(1 - \kappa) \,[\ln(1 - \kappa)]^2}, & \kappa \to 1\,,
\end{cases}  
\end{equation}}
implying that \(H(\kappa)\) vanishes rapidly as \(\kappa\to 0\) and diverges in an integrable way as \(\kappa\to 1^-\).

The variance for arbitrary resetting protocol $h(\tau)$ is given in Eq. (\ref{def_Var2}). For the Poissonian case it reads
\bea \label{def_Var2.1}
{\rm Var}(N_L) =  N\,r \int_0^\infty d\tau \, e^{-r\tau}  q(\tau) [1 - q(\tau)] \;,
\eea
where $q(\tau) =  {\rm erf}(L/\sqrt{4 D \tau})$. Here we study this variance as a function of $L$. Indeed, the size dependence of the variance for the Poissonian resetting case was not analysed in Ref. \cite{gas_bosons}. We show here that it has an interesting non-monotonic behavior as a function of $L$. The result in Eq. (\ref{def_Var2.1}) can be expressed in the scaling form
\bea \label{var_scaling}
{\rm Var}(N_L) = N\, V\left( \sigma = \frac{r\, L^2}{4 D}\right) \;,
\eea
where the scaling function $V(\sigma)$ is given by
\bea \label{V_gamma0}
V(\sigma) = 2 \sigma \, \int_0^\infty \frac{du}{u^3}\, e^{-\frac{z}{u^2}}\, {\rm erf}(u){\rm erfc}(u) \;.
\eea
The scaling function $V(\sigma)$ is plotted as a function of $\sigma$ in Fig. \ref{fig_V}. It has a nonmonotonic behavior, namely
\bea \label{asympt_V}
V(\sigma) \approx
\begin{cases}
& 2 \sqrt{\sigma} \quad, \quad \;\, \sigma \to 0 \\
& \\
& e^{- 2 \sqrt{\sigma}} \quad, \quad \sigma \to \infty \;.
\end{cases}
\eea 
Thus the variance decays exponentially as $e^{-L/l}$ where $l = \sqrt{D/r}$ denotes the typical distance that a particle travels between two successive resettings. Such a nonmonotonic behavior has also been found for the one-dimensional log-gas associated to Gaussian random matrix ensembles~\cite{marino2014phase,marino2016number}, where the eigenvalues are supported over a finite region. There, as a function of $L$, the variance increases logarithmically with $L$ and then drops extremely sharply when $L$ exceeds the support of the eigenvalues, indicating that the eigenvalues form a rigid and compact gas over the support. Any such one-dimensional gas where the number variance grows slower than $L$ is called ``hyperuniform'' \cite{torquato2018hyperuniform}. In our case, the variance initially grows as $L$ (since $V(\sigma) \sim \sqrt{\sigma}$) for small $L \ll l$, indicating that the gas is not hyperuniform for small $L$. However, for large $L$, it decays exponentially with $L$, signalling a hyperunifiorm behavior. Indeed, the exponential decay for large $L$ indicates that, even though the particles are not supported over a finite interval, the gas is still somewhat ``rigid''.

\subsection{Power-law (Pareto) resetting protocol}
Next, we examine the Pareto law for the interval (between resettings) distribution
\begin{equation}\label{eq:powlawdef}
    \psi(\tau)=\frac{r\beta}{(r\tau)^{1+\beta}} \quad, \quad {\rm with} \quad \tau\in [1/r,\infty)
\end{equation}
defined for \(\beta>0\). To ensure a stationary state (i.e. finite \(\langle \tau\rangle\)), we restrict the discussion to \(\beta>1\). In this case, the lower cutoff $\tau = 1/r$ in $\psi(\tau)$ indicates that there is never a resetting interval whose duration is less that $1/r$.
This means that, once a resetting occurs, the next resetting event will occur only after a minimum time $1/r$. At the end of this refractory period of duration $1/r$, the mean number of particles in the interval $[-L,L]$ is given, from Eq. (\ref{eq:fulcstat}),~by 
\bea \label{NL1or}
\langle N_L(\tau = {1}/{r})\rangle  = N\, q(\tau = 1/r) = N\, {\rm erf}(\sqrt{\sigma}) \quad, \quad \sigma = \frac{r\,L^2}{4\,D} \;.
\eea
Thus the associated mean fraction of particles in $[-L,L]$ at the end of this refractory period is given by
\bea \label{def_kappastar}
\kappa^* =  {\rm erf}(\sqrt{\sigma}) \;.
\eea
We will see later that this value $\kappa^*$ plays a crucial role in the analysis of the FCS.

Using the definition \eqref{eq:h(t)} we first compute the normalization
\bea \label{Psi_Pareto}
\int_0^{\infty} dt\, \Psi(t)=\frac{\beta}{\beta-1}\frac{1}{r}\,,
\eea
which leads to
\begin{equation}\label{eq:h_plaw}
    h(\tau) =
    \begin{cases}
    \displaystyle 
    \frac{\beta-1}{\beta}\,r, & \tau < 1/r\,, \\[1.0em]
    \displaystyle
    \frac{\beta-1}{\beta}\,r\,(r\tau)^{-\beta}, & \tau > 1/r\,.
    \end{cases}
\end{equation}
We now present our results for the different observables, as in the Poissonian case in the previous subsection.

\begin{figure}[t]
    \centering
    \includegraphics[width=0.45\textwidth]{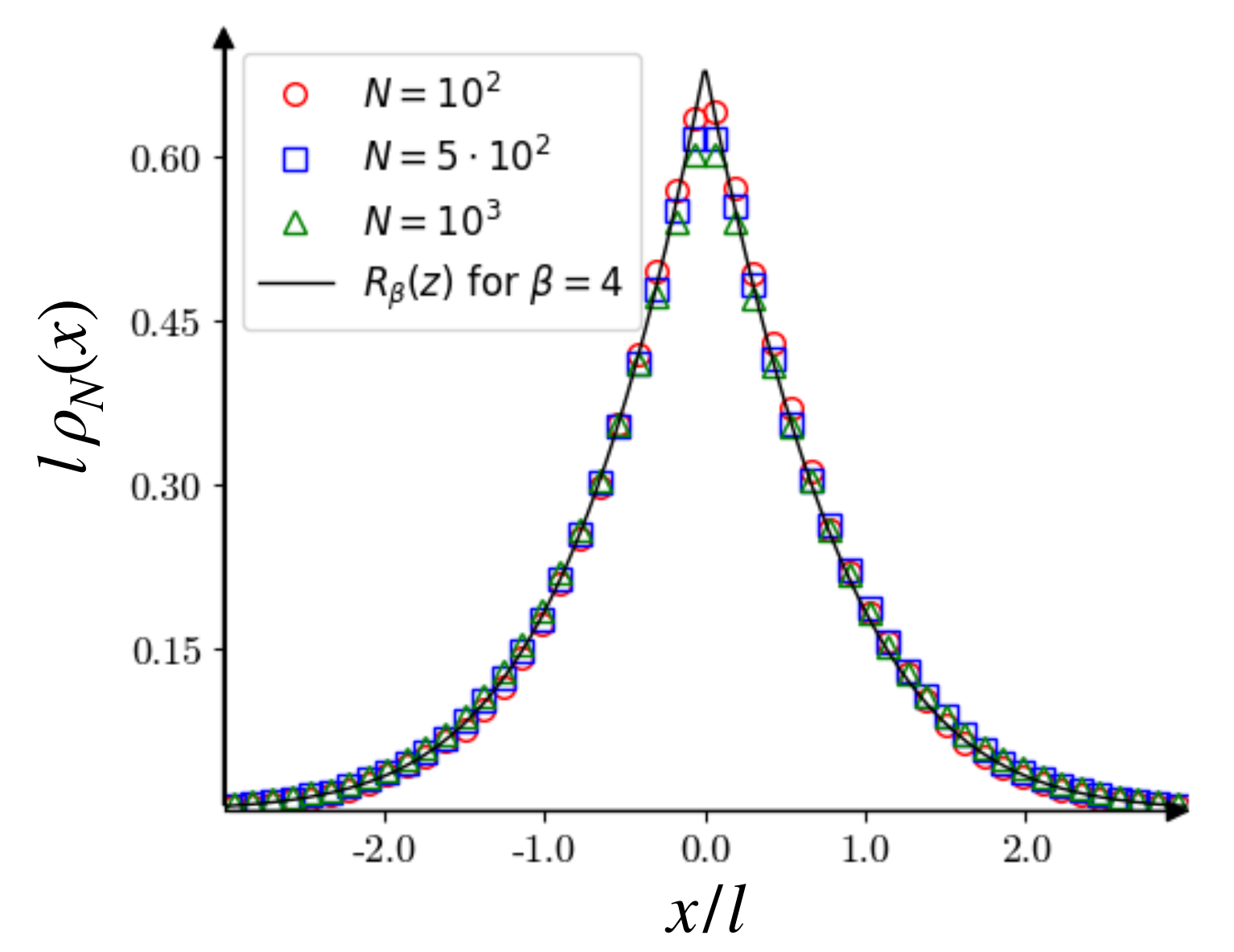}
    \caption{Plot of the density profile $\rho_N(x) = \frac{1}{l}\, R_\beta\left( \frac{x}{l} \right)$. 
    The scaling function $R_\beta(z)$ is given in~\eqref{eq:density_R(z)}, and $l = \sqrt{\frac{D}{r}}$. 
    The solid black curve shows the theoretical prediction for the resetting protocol 
    $\psi(t) = \frac{r\beta}{(rt)^{1+\beta}}$ with $t \in [1/r, \infty)$ and $\beta = 4$. 
    The symbols indicate simulation results for different values of $N$. 
    Simulations were performed with $r = 1$ and $D = \frac{1}{2}$. The plots for different values of $N$ clearly show that $\rho_N(x)$ is independent of $N$.}
    \label{fig:R_pl}
\end{figure}

\vspace*{0.5cm}
\noindent{\it $\bullet$ Average density and correlations.} Inserting Eq. \eqref{eq:h_plaw} in Eq. \eqref{eq:rhodef} we can write the density \(\rho_N(x)\) in the scaling form 
\begin{equation}\label{eq:density_scalingform}
    \rho_N(x)=\frac{1}{l}\, R_\beta\!\left( \frac{x}{l}\right)\,,
\end{equation}
where the length scale $l=\sqrt{\frac{D}{r}}$ is the same as in the Poissonian case and where the scaling function \(R_\beta(z)\) is given by
\begin{equation}\label{eq:density_R(z)}
    R_\beta(z)=\frac{z}{4\sqrt{\pi}}\frac{(\beta-1)}{\beta} \left[ \Gamma\left(-\frac{1}{2},\frac{z^2}{4}\right)+\left(\frac{2}{z}\right)^{2\beta}\gamma\left(\beta-\frac{1}{2},\frac{z^2}{4}\right) \right] \,.
\end{equation}
The function $\Gamma(s,z)$ and $\gamma(s,z)$ are respectively the upper and lower incomplete gamma functions, defined as $\Gamma(s,z)= \int_0^{z}dy\, y^{s-1}\, e^{-y}$ and $\gamma(s,z)= \int_z^{\infty}dy\, y^{s-1}\, e^{-y}$. The resulting density profile, shown in Figure \ref{fig:R_pl}, exhibits the following asymptotic behaviors
\begin{equation}
R_\beta(z)\approx
    \begin{cases}
    A_1-A_2\,\lvert z \rvert, & z \to 0\,, \\[1mm]
    \displaystyle \frac{A_3}{z^{2\beta-1}}, & z \to \infty\,.
    \end{cases}
\end{equation}
where $A_1, A_2$ and $A_3$ are positive constants.
This result is particularly interesting because it tells us that the power-law resetting distribution leads to a power-law tail in the particle density, in contrast with the exponential decay found in the Poissonian case. Although the scale factor $l = \sqrt{D/r}$ remains unchanged, longer intervals between successive resettings occur with a higher probability, allowing particles to diffuse further from the origin. This results in a fat-tailed average density.

In the NESS, correlations are again strong. Indeed, substituting $h(\tau)$ from (\ref{eq:h_plaw}) in Eq. (\ref{C2_inf}), we get 
\begin{equation}
  {\cal C}_2(t \to \infty) =   \langle x_i^2 x_j^2\rangle_c = \langle x_i^2 x_j^2\rangle - \langle x_i^2\rangle \langle x_j^2\rangle = \frac{4D^2}{r^2} \frac{(\beta-1)(\beta^2-4\beta+7)}{12(\beta-3)(\beta-2)^2}  \quad \forall \, i, j\,,
\end{equation}
which is defined for \(\beta>3\), i.e. when the resetting time variance is finite.

\begin{figure}[t]
    \centering
    \includegraphics[width=0.4\textwidth]{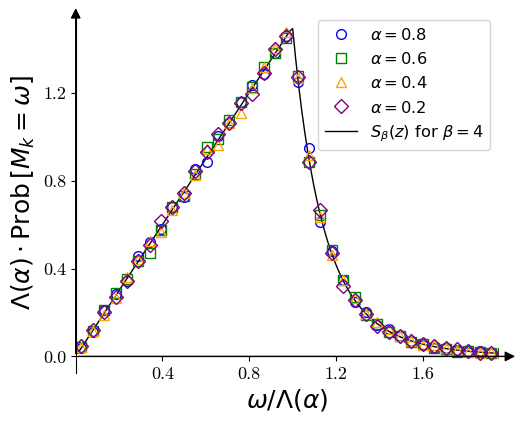}
    \caption{Plot of the distribution of the $k^{th}$ rightmost particle $M_k$ (order statistics). The distribution follows the scaling form 
    $\text{Prob} [ M_k = \omega ]\approx \frac{1}{\Lambda(\alpha)}\, S_\beta\left( \frac{\omega} {\Lambda(\alpha)} \right)$ for large $N$, 
    where the scaling function $S_\beta(z)$ is given in~\eqref{eq:S_plaw}, and the scaling parameter $\Lambda(\alpha)$ is defined in~\eqref{eq:lambd_alpha_def}. 
    The solid black curve represents the theoretical prediction for the resetting protocol 
    $\psi(t) = \frac{r\beta}{(rt)^{1+\beta}}$ with $t \in [1/r, \infty)$ and $\beta = 4$. 
    The symbols represent results from simulations performed with $N = 10^6$ particles, $r = 1$, and $D = \frac{1}{2}$, for different values of $\alpha = \frac{k}{N}$. The scaling function $S_\beta(z)$ is clearly independent of $\alpha$.}
    \label{fig:S_pl}
\end{figure}

\vspace*{0.5cm}
\noindent{\it $\bullet$ Order statistics.}
The statistics of the extremes are modified as well. Setting $k = \alpha N$ with $0<\alpha<1/2$ as before, one finds
\begin{equation}\label{eq:scalingform_Mk_sez_plaw}
    \text{Prob} [ M_k = \omega] \approx \frac{1}{\Lambda(\alpha)} S_\beta \!\left(  \frac{\omega}{\Lambda(\alpha)} \right)\,,
\end{equation}
where the scaling function $S_\beta(z)$ is given by 
\begin{equation}
S_\beta(z)=
\begin{cases}\label{eq:S_plaw}
    2\frac{\beta-1}{\beta}\,z, & z<1\,, \\[1.0em]
    2\frac{\beta-1}{\beta}\,\frac{1}{z^{2\beta-1}}, & z>1\,.
\end{cases}
\end{equation}
For a plot of this scaling function and also comparison to numerical simulations, see Fig.~\ref{fig:S_pl}. 
The scaling factor \(\Lambda(\alpha)\) remains the same as in the Poissonian case and is given in Eq. (\ref{eq:lambd_alpha_def}). Thus, as in the Poissonian case, the dependence on $k$ of the distribution of $M_k$ is only the scale factor $\Lambda(\alpha)$, but the scaling function $S_\beta(z)$ is universal, i.e., independent of $k$. However, as in the Poissonian case, exactly at $\alpha = 1/2$, the scaling form is again given by
\bea \label{Mk_middle_pareto}
 \text{Prob} [M_{k=N/2} = \omega ] \approx \sqrt{\tilde b\,N}\, R_\beta \left( \sqrt{\tilde b\, N}\, \omega \right) \;,
\eea
where $\tilde b$ is just an unimportant scale factor (which has a cumbersome expression not given here) and $R_\beta(z)$ is exactly the scaling function associated with the average density in Eq. (\ref{eq:density_R(z)}).

The average density \(\rho_N(x)\) is also the one-point marginal of the particle positions, and one may compare the above EVS, given by \eqref{eq:scalingform_Mk_sez_plaw} for $k=O(1)$, with that obtained by sampling \(N\) IID random variables from \(\rho_N(x)\) in Eq. (\ref{eq:density_scalingform}). In this latter case, the limit \(N\to \infty\) gives a Fr\'echet distribution $\tilde{S}_\beta(z)=\frac{\alpha}{z^{\alpha+1}}e^{-z^{-\alpha}}$ with parameter \(\alpha=2\beta-2\) (see Appendix \ref{app:iidvar}). It behaves as
\begin{equation}
\tilde{S}_\beta(z) \approx
    \begin{cases}
        \exp\!\big[-z^{-(2\beta-2)}\big], & z \to 0\,, \\[1mm]
        \displaystyle \frac{1}{z^{2\beta-1}}, & z \to \infty\,.
    \end{cases}
\end{equation}
Thus, the resetting protocol increases the probability mass near the origin while preserving the power-law tail for large \(z\).

\begin{figure}[t]
    \centering
    \includegraphics[width=0.4\textwidth]{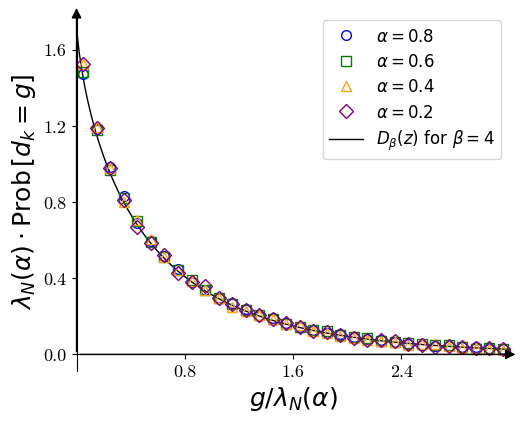}
    \caption{Plot of the distribution of the spacing $d_k=M_k-M_{k+1}$ between the $k^{th}$ and $(k+1)-$th rightmost particles. The distribution follows the scaling form $\text{Prob}[d_k = g] \approx \frac{1}{\lambda_N(\alpha)} D_\beta\left(\frac{g}{\lambda_N(\alpha)} \right)$ for large $N$, where the scaling function $D_\beta(z)$ is given by \eqref{eq:D_plaw_eq}, and the scaling parameter $\lambda_N(\alpha)$ is defined in \eqref{eq:lambdaN_definition}.
    The solid black curve represents the theoretical prediction for the resetting protocol $\psi(t)=\frac{r\beta}{(rt)^{1+\beta}}$ with \(t\in [1/r,\infty)\) and $\beta=4$. The symbols represent results from simulations performed with $N = 10^6$ particles, $r = 1$, and $D = \frac{1}{2}$, for different values of $\alpha = \frac{k}{N}$.}
    \label{fig:D_pl}
\end{figure}

\vspace*{0.5cm}
\noindent{\it $\bullet$ Gap statistics.}
The gap distribution $\text{Prob}[d_k = g]$ can be computed exactly from Eq. (\ref{eq:d_k_p(omega)}). It again exhibits a scaling form 
\begin{equation}\label{ea:dk_generalforh}
\text{Prob}[d_k = g] = \frac{1}{\lambda_N(\alpha)}\, D_\beta \!\left(\frac{g}{\lambda_N(\alpha)} \right)\,,
\end{equation}
where the scale factor $\lambda_N(\alpha)$ is given in Eq. (\ref{eq:lambdaN_definition}) and the scaling function $D_\beta(z)$ reads
\begin{equation}\label{eq:D_plaw_eq}
    D_\beta(z)=2\frac{(\beta-1)}{\beta}\left[e^{-z} -z\,\Gamma(0,z)+z^{2\beta-1}\left\{\Gamma(2\beta-1)-\gamma\left(2\beta-1,z\right) \right\}\right]\,.
\end{equation}
The function $D_\beta(z)$ is plotted, along with comparison with numerical simulations, in Fig.~\ref{fig:D_pl}. 
The scaling function $D_\beta(z)$ has the asymptotic behaviors
\begin{equation}\label{eq:D(z)_powlaw}
D_\beta(z)\approx
    \begin{cases}
        A_4+A_5z\ln z, & z \to 0\,, \\[1mm]
        & \\
        \displaystyle \frac{A_6}{z^{2\beta-1}}, & z \to \infty\,,
    \end{cases}
\end{equation}
where $A_4$, $A_5$ and $A_6$ are positive constants. While for $z \to 0$, the scaling function $D_\beta(z)$ approaches a constant as in the Poissonian case, for large $z$, it exhibits a power-law tail (as opposed to the stretched exponential tail found in the Poissonian case (\ref{D_asympt_Pois})), indicating the presence of large gaps with higher probability in the stationary state.

It is also interesting to compare this distribution with what we would have obtained sampling $N$ IID random variables from the average density $\rho_N(x)$ in Eq. (\ref{eq:density_scalingform}). For the Fr\'echet class, as in this case, the asymptotic scaling function $\tilde{D}_\beta(z)$ for the distribution of the first gap $d_1$ is such that (cf. Appendix~\ref{app:iidvar}):
\begin{equation}\label{eq:d_k_frechet}
\tilde{D}_\beta(z)\approx
    \begin{cases}
        A_7-A_8z & z \to 0 \;,
    \\
    & \\
        \frac{\alpha}{(k-1)!} \frac{1}{z^{1+\alpha}} & z \to \infty \;,
    \end{cases}
\end{equation}
where $A_7$ and $A_8$ are positive constants and
 $\alpha=2\beta-2$. Thus, we have a slightly different behavior for small $z$, but the tails at $z \to \infty$ of \eqref{eq:D(z)_powlaw} and \eqref{eq:d_k_frechet} are the same.\\

\begin{figure}[t]
    \centering
    \includegraphics[width=0.9\textwidth]{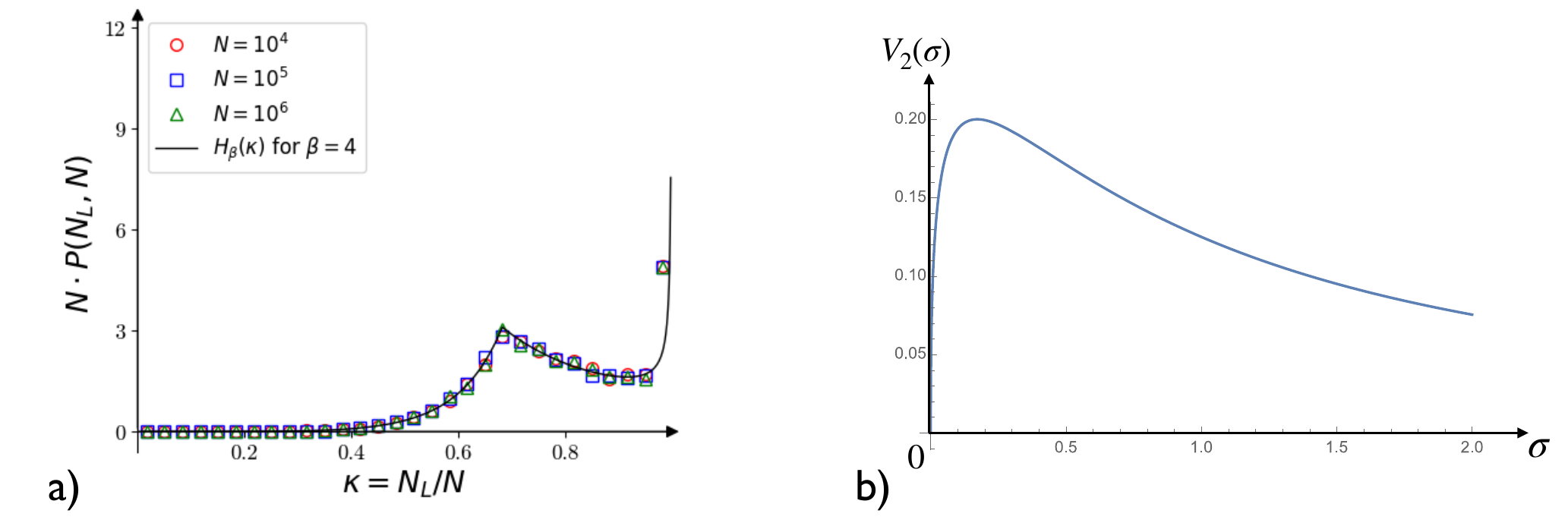}
     \caption{{\bf a)}: Plot of the full counting statistics given by the distribution $P(N_L,N)\approx\frac{1}{N}H_\beta\left( \frac{N_L}{N} \right)$. The scaling function $H(\kappa)$ is given by \eqref{eq:H_plaw_explicit}. The solid black curve shows the theoretical prediction for the resetting protocol $\psi(t)=\frac{r\beta}{(rt)^{1+\beta}}$ with \(t\in [1/r,\infty)\) and $\beta=4$. The symbols indicate simulation results for different values of $N$. The simulations were done with $r=1$ and $D=\frac{1}{2}$. Note that the cusp occurs at $\kappa = \kappa^*  = {\rm erf}(\sqrt{\sigma})$. {\bf b)}: Plot of the function $V_2(\sigma)$ vs $\sigma$, as given in Eq. (\ref{V_beta}), setting $\beta=2$.}\label{FCS_Pareto}
\end{figure}

\vspace*{0.5cm}
\noindent{\it $\bullet$ Full counting statistics and variance.}
Similarly, the full counting statistics (FCS) admits, for large $N$, the scaling form $P(N_L,N)=\frac{1}{N}H_\beta\left( \frac{N_L}{N} \right)$ in Eq.~(\ref{eq:fullcount_tot}) where the scaling function $H_\beta(\kappa)$ can be computed from the integral in Eq.~(\ref{eq:fullcount_tot}). We get
\begin{equation}\label{eq:H_plaw_explicit}
    H_\beta(\kappa)=
    \begin{cases}
        \sqrt{\pi}\,\sigma^{1-\beta}\left( \frac{\beta-1}{\beta}\right) \left[u(\kappa)\right]^{2\beta-3}e^{u^2(\kappa)}, & \kappa<\kappa^*=\operatorname{erf}\!\left( \sqrt{\sigma} \right)\,,\\[2mm]
        \\
        \sigma \sqrt{\pi}\left( \frac{\beta-1}{\beta}\right) \left[u(\kappa)\right]^{-3}e^{u^2(\kappa)}, & \kappa>\kappa^* = \operatorname{erf}\!\left( \sqrt{\sigma} \right)\,, 
    \end{cases}
\end{equation}
where $\kappa=\frac{N_L}{N}$ represents the fraction of particles in $[-L,L]$. Here  \(\sigma=\frac{rL^2}{4D}\), as before, and 
\(u(\kappa)=\operatorname{erf}^{-1}(\kappa)\). This function is highly singular at $\kappa = \kappa^* = {\erf}(\sqrt{\sigma})$. 
For a plot of it, and comparison to numerical simulations, see Fig.~\ref{FCS_Pareto}~a). The function $H_\beta(\kappa)$ has the following asymptotic behaviors
\blue{\bea \label{asympt_Hbeta}
H_{\beta}(\kappa) \approx
\begin{cases}
&\sqrt{\pi}\,\sigma^{1-\beta}\left( \frac{\beta-1}{\beta}\right)\left(\frac{\sqrt{\pi}}{2}\,\kappa \right)^{2\beta-3} \quad, \quad \kappa \to 0 \\
& \\
&\sigma \left( \frac{\beta-1}{\beta}\right) \frac{1}{(1-\kappa)\, \left|\ln (1-\kappa)\right|^2} \quad, \quad \kappa \to 1 \;.
\end{cases}
\eea}

Comparing with the Poissonian case in Eq. (\ref{eq:asympt_H_poiss}) we find that, while the scaling function $H_\beta(\kappa)$ has a similar integrable divergence as $\kappa \to 1$, the behavior near $\kappa \to 0$ is drastically different. As $\kappa \to 0$, the scaling function $H_\beta(\kappa)$ vanishes much slower as a power law, as opposed to the much faster essential singular vanishing for the Poissonian case.  
Thus under the power-law resetting protocol, the probability of having $\kappa$ close to $0$, i.e., to have a ``hole'' devoid of particles near the origin is higher compared to the Poissonian case. Physically, this can be understood from the fact that resetting happens much less frequently in the power law protocol and hence the particles have more time to diffuse away from the origin. 

In addition to these asymptotic behaviors as $\kappa \to 0$ and $\kappa \to 1$, we note from Fig. \ref{FCS_Pareto} a) that the function $H_\beta(\kappa)$ is highly singular at $\kappa = \kappa^* = {\rm erf}(\sqrt{\sigma})$, which is also evident from the explicit expression in Eq. (\ref{eq:H_plaw_explicit}). As $\kappa \to \kappa^*$ from left and right, while the scaling function $H_\beta(\kappa)$ remains continuous, its derivative undergoes a discontinuous jump at $\kappa = \kappa^*$. Indeed we find 
\bea \label{disc_H}
H_{\beta}(\kappa = \kappa^*) &=& \frac{\sqrt{\pi} (\beta -1)}{\beta} \frac{e^{\sigma}}{\sqrt{\sigma}}  \\
H'_{\beta}(\kappa = {\kappa^*}^+) - H'_{\beta}(\kappa = {\kappa^*}^-)&=& -\frac{(\beta-1)}{\sigma} \pi e^{2 \sigma} \;.
\eea 
The existence of such a cusp in the FCS at an intermediate value $\kappa = \kappa^*$ is rather unusual. In this case, this cusp can be  
traced back to the fact that the interval distribution $\psi(\tau) = 0$ for $\tau \in [0,1/r]$, i.e., the existence of a minimal interval duration $\tau = 1/r$.

Finally we turn to the variance of $N_L$, given in Eq. \eqref{def_Var2}. Substituting $h(\tau)$ given in Eq. (\ref{eq:h_plaw}) in  \eqref{def_Var2}, the variance can again be written in the scaling form
\bea \label{var_scaling_beta}
{\rm Var}(N_L) = N\, V_\beta\left( \sigma = \frac{r\, L^2}{4 D}\right) \;,
\eea
where the scaling function $V_\beta(\sigma)$ is now given by
\bea \label{V_beta}
V_\beta(\sigma) = 2 \sigma \frac{(\beta-1)}{\beta} \left[ \frac{1}{\sigma^\beta} \int_0^{\sqrt{\sigma}} du\, u^{2\beta - 3} {\rm erf}(u){\rm erfc}(u) + \int_{\sqrt{\sigma}}^\infty du\, u^{-3}  {\rm erf}(u){\rm erfc}(u) \right] \;,
\eea
where we recall that $\beta > 1$. For general $\beta >1$, it is hard to perform the integrals explicitly. However one can derive the asymptotics for large and small $\sigma$ and they are given by
\bea \label{asympt_V_beta}
V_\beta(\sigma) \approx
\begin{cases}
& \frac{8(\beta-1)}{\sqrt{\pi}(2 \beta - 1)}\sqrt{\sigma} \quad, \quad \;\, \sigma \to 0 \\
& \\
& c_\beta\,\sigma^{1-\beta} \quad, \quad \quad \quad \; \sigma \to \infty \;,
\end{cases}
\eea 
where the prefactor $c_\beta$ is given by
\bea \label{cb}
c_\beta = 2 \frac{(\beta -1)}{\beta} \int_0^{\infty} du u^{2\beta-3}\, {\rm erf}(u) {\rm erfc}(u) \;.
\eea
Thus $V_\beta(\sigma)$, as a function of $\sigma$, is again non-monotonic for any $\beta > 1$. For $\beta=2$ one can calculate explicitly $V_2(\sigma)$ for all $\sigma$ and plot it as shown in Fig. \ref{FCS_Pareto} b).

\subsection{Bounded resetting protocol}

\begin{figure}[t]
    \centering
    \includegraphics[width=0.45\textwidth]{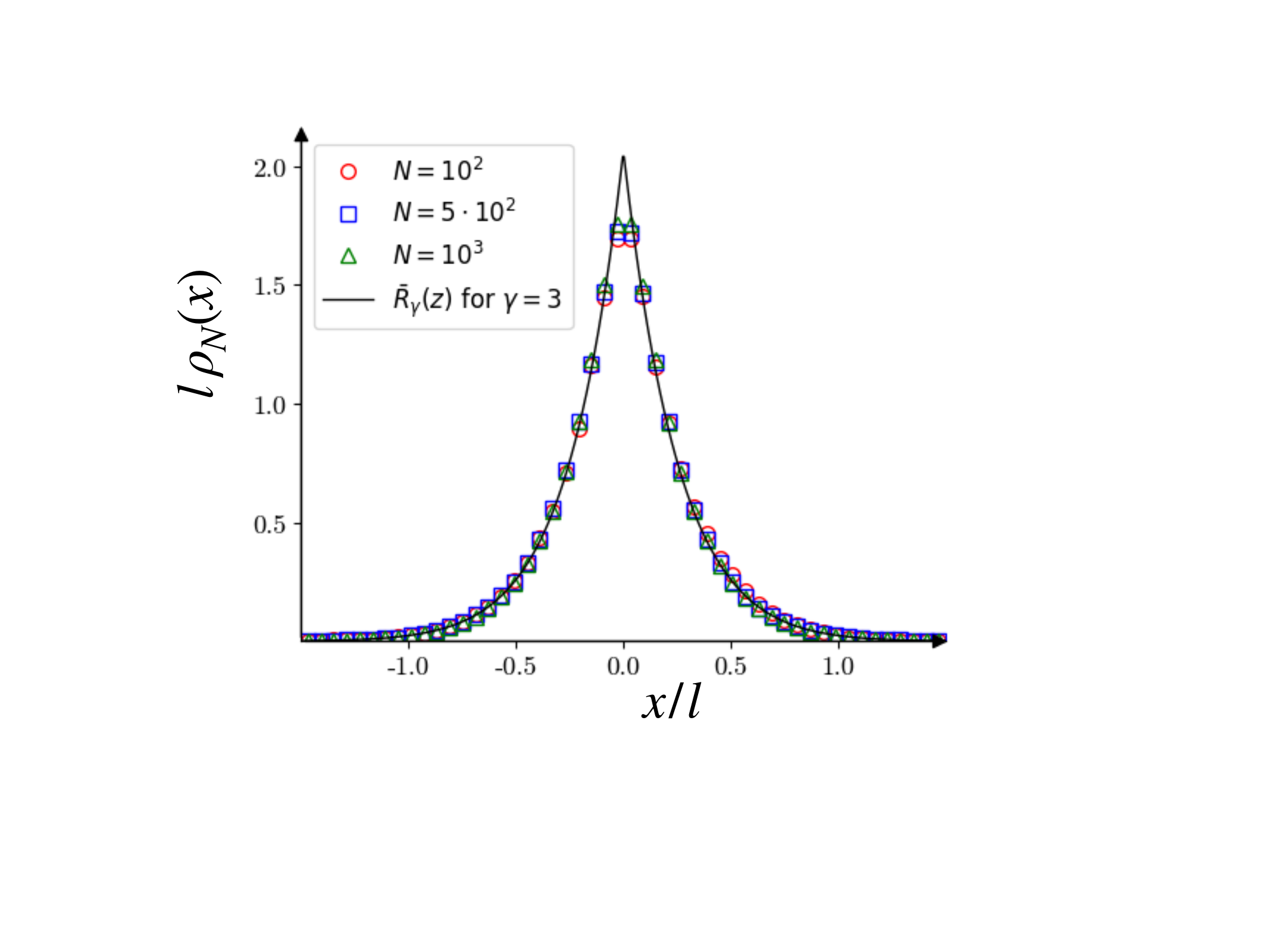}
    \caption{Plot of the density profile $\rho_N(x) = \frac{1}{l}\, \bar{R}_\gamma\left( \frac{x}{l} \right)$. 
    The scaling function $\bar{R}_\gamma(z)$ is given in~\eqref{eq:R_weib_explict}, and $l = \sqrt{\frac{D}{r}}$. 
    The solid black curve shows the theoretical prediction for the resetting protocol 
    $\psi(\tau) = \frac{\gamma r}{T^{\gamma}} \left(T - r\tau \right)^{\gamma - 1}$ 
    with $\tau \in [0, T/r]$, $\gamma = 3$, and $T = 1$. 
    The symbols indicate simulation results for different values of $N$ (clearly the density is independent of $N$). 
    Simulations were performed with $r = 1$ and $D = \frac{1}{2}$.}
    \label{fig:R_w}
\end{figure}
Finally, we consider a resetting distribution with a finite support, defined as
\begin{equation}\label{eq:weibull_defpsi}
    \psi(\tau)=\frac{\gamma r}{T^{\gamma}} \left(T-r\tau \right)^{\gamma-1} \quad, \quad {\rm for} \quad \tau \in[0,T/r] \;,
\end{equation}
defined for any \(\gamma>0\). Note that $T$ in Eq. (\ref{eq:weibull_defpsi}) is a dimensionless positive number. 
Since the support of the distribution is upper-bounded by \(\tau=T/r\), the maximum time between resets is \(T/r\). Using the definition
\[
h(\tau) = \frac{\Psi(\tau)}{\int_0^{\infty} dt\, \Psi(t)}\,,
\]
one finds
\begin{equation} \label{def_h_weib}
    h(\tau)=
    \begin{cases}
        \displaystyle \frac{\gamma+1}{T^{\gamma+1}}\, r \, \left(T-r\tau \right)^{\gamma}, & \tau \in[0,T/r)\,, \\[1mm]
        & \\
        0, & \tau \geq T/r\,.
    \end{cases}
\end{equation}
We compute below the different observables as before.

\vspace*{0.5cm}
\noindent{\it $\bullet$ Average density and correlations.}
Plugging in this expression for $h(\tau)$ in Eq. (\ref{eq:rhodef}) for the average density, we get 
\begin{equation} \label{rho_weib}
    \rho_N(x)=\frac{1}{l}\, \bar{R}_\gamma\!\left( \frac{x}{l}\right)\,,
\end{equation}
with
\begin{equation}\label{eq:R_weib_explict}
    \bar{R}_\gamma\!\left( z\right)=\frac{\gamma+1}{T^{\gamma+1}\sqrt{\pi}}\int_0^{\sqrt{T}}du\, \left( T-u^2\right)^{\gamma} \exp\!\left[ -\left(\frac{z}{2}\right)^2\frac{1}{u^2} \right]\,,
\end{equation}
where \(l=\sqrt{\frac{D}{r}}\) as in the previous cases. For this bounded distribution $\psi(\tau)$, we choose the notation $\bar{R}_\gamma(z)$ for the scaling function to distinguish it from the Poissonian and the Pareto cases discussed before. This scaling function $\bar{R}_\gamma(z)$ has the asymptotic behaviors
\begin{equation} \label{Rg_asympt}
    \bar{R}_\gamma\left( z\right)\approx
    \begin{cases}
        B_1-B_2\, \lvert z \rvert, & z \to 0\,, \\[1mm]
        & \\
        \blue{B_3 \, z^{-2(\gamma+1)} \, e^{-\frac{z^2}{4T}}}, & z \to \infty\,,
    \end{cases}
\end{equation}
where $B_1, B_2$ and $B_3$ are computable positive constants. This scaling function $\bar{R}_\gamma(z)$ is plotted in Fig.~\ref{fig:R_w} and is compared to numerical simulations. The faster decay at large distances, compared to the Poissonian case [see Eq.~(\ref{Rofz})], is a direct consequence of the finite cut-off in the resetting time distribution.

As in previous cases, the resetting mechanism generates strong correlations between particles. Specifically, substituting $h(\tau)$ from Eq. (\ref{def_h_weib}) in (\ref{C2_inf}), the second-order connected two-point correlator in the NESS reads
\begin{equation} \label{C2gamma}
  {\cal C}_2(t \to \infty) =\langle x_i^2 x_j^2\rangle_c = \langle x_i^2 x_j^2\rangle - \langle x_i^2\rangle \langle x_j^2\rangle = \frac{4D^2}{r^2}\frac{\gamma + 1}{(\gamma + 2)^2 (\gamma + 3)} \;,
\end{equation}
which is smaller than that in Eq.~\eqref{eq:correlator_poissonian} for any $\gamma$, but still indicates strong correlations.

\begin{figure}[t]
    \centering
    \includegraphics[width=0.4\textwidth]{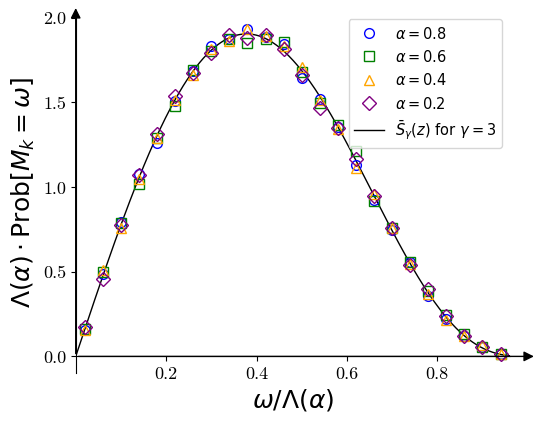}
    \caption{Plot of the distribution of the $k^{th}$ rightmost particle $M_k$ (order statistics). The distribution follows the scaling form 
    $\text{Prob} [ M_k = \omega ] \approx \frac{1}{\Lambda(\alpha)}\, \bar{S}_\gamma\left( \frac{\omega} {\Lambda(\alpha)} \right)$ for large $N$, 
    where the scaling function $\bar{S}_\gamma(z)$ is given in~\eqref{eq:S_weib_exact}, and the scale factor $\Lambda(\alpha)$ is defined in~\eqref{eq:lambd_alpha_def}. 
    The solid black curve represents the theoretical prediction for the resetting protocol $\psi(\tau)=\frac{\gamma r}{T^{\gamma}} \left(T-r\tau \right)^{\gamma-1}$ for \(\tau \in[0,T/r]\), $\gamma=3$ and $T=1$. 
    The symbols represent results from simulations performed with $N = 10^6$ particles, $r = 1$, and $D = \frac{1}{2}$, for different values of $\alpha = \frac{k}{N}$.}
    \label{fig:S_w}
\end{figure}

\vspace*{0.5cm}
\noindent{\it $\bullet$ Order statistics.} Having characterized the correlations, we now turn to the order statistics of the system by studying the $k^{th}$ maximum $M_k$ among the particles' positions. As in the Poissonian and power-law cases studied before, one can express the distribution of $M_k$ in the large $N$ limit via the scaling form 
\begin{equation}\label{eq:scalingform_Mk_sez_weib}
    \text{Prob} [ M_k = \omega]\approx \frac{1}{\Lambda(\alpha)} \bar{S}_\gamma \!\left(  \frac{\omega}{\Lambda(\alpha)} \right)\,,
\end{equation}
where $\Lambda(\alpha)$ is given in Eq. (\ref{eq:lambd_alpha_def}) and the scaling function $\bar{S}_\gamma(z)$ is given by
\begin{equation}\label{eq:S_weib_exact}
   \bar{S}_\gamma(z)=
    \begin{cases}
        \displaystyle \frac{2(\gamma+1)}{T^{\gamma+1}}\,z\,(T-z^2)^{\gamma}, & z<\sqrt{T}\,, \\[1mm]
        0, & z \geq \sqrt{T}\,.
    \end{cases}
\end{equation}
This scaling function, along with numerical data for various values of $\alpha=\frac{k}{N}$, is shown in Fig.~\ref{fig:S_w}. Exactly for $\alpha = 1/2$, as in the previous cases, the scaling behavior of the PDF of $M_{k=N/2}$, up to a trivial scale factor, is again given by the same scaling function $\bar{R}_\gamma(z)$ associated to the global average density given in Eq. (\ref{eq:R_weib_explict}).  

It is interesting to compare this result in Eq. (\ref{eq:S_weib_exact}) for the ordered maxima to that of $N$ IID variables, each drawn from a PDF equalling the average density $\rho_N(x)$ in Eq. (\ref{rho_weib}). Since this PDF decays as a Gaussian for large argument (see Eq. (\ref{Rg_asympt})), one would expect~\cite{StatisticsofExtremesandRecordsinRandomSequence} (see also Appendix~\ref{app:iidvar}) that the scaled distribution of the $k$-th maximum should have a generalized Gumbel given in Eq. (\ref{kGumbel}). In contrast to this expectation, the result in Eq. (\ref{eq:S_weib_exact}) shows that the support of the distribution is instead upper-bounded and very different from a generalized Gumbel. This reflects the effects of strong correlations between the particles.

\begin{figure}[t]
    \centering
    \includegraphics[width=0.4\textwidth]{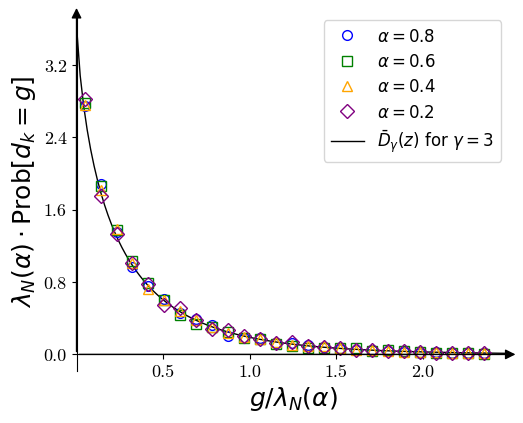}
    \caption{Plot of the distribution of the spacing $d_k=M_k-M_{k+1}$ between the $k^{th}$ and $(k+1)-$th rightmost particles. The distribution follows the scaling form $\text{Prob}[d_k = g] \approx \frac{1}{\lambda_N(\alpha)} \bar{D}_\gamma\left(\frac{g}{\lambda_N(\alpha)} \right)$ for large $N$, where the scaling function $\bar{D}_\gamma(z)$ is given by \eqref{eq:D_weib_explicit}, and the scale factor $\lambda_N(\alpha)$ is defined in \eqref{eq:lambdaN_definition}.
    The solid black curve represents the theoretical prediction for the resetting protocol $\psi(\tau)=\frac{\gamma r}{T^{\gamma}} \left(T-r\tau \right)^{\gamma-1}$ for \(\tau \in[0,T/r]\), $\gamma=3$ and $T=1$. The symbols represent results from simulations performed with $N = 10^6$ particles, $r = 1$, and $D = \frac{1}{2}$, for different values of $\alpha = \frac{k}{N}$.}
    \label{fig:D_w}
\end{figure}

\vspace*{0.5cm}
\noindent{\it $\bullet$ Gap distribution.}
Following exactly the same steps as before, the large $N$ behavior of the spacing distribution can be written as
\begin{equation}\label{ea:dkda_integr}
\text{Prob}[d_k = g] \approx \frac{1}{\lambda_N(\alpha)}\, \bar{D}_\gamma \!\left(\frac{g}{\lambda_N(\alpha)} \right)\,,
\end{equation}
where $\lambda_N(\alpha)$ is given in Eq. (\ref{eq:lambdaN_definition}) and the scaling function $\bar{D}_\gamma(z)$ reads
\begin{equation}\label{eq:D_weib_explicit}
    \bar{D}_\gamma(z)=2\frac{(\gamma+1)}{T^{\gamma+1}}\int_0^{\sqrt{T}}du\, \left( T-u^2\right)^{\gamma} \exp\!\left(-\frac{z}{u}\right)\,.
\end{equation}
A comparison between the theoretical prediction and simulations is shown in Fig.~\ref{fig:D_w} and its asymptotic behaviors are
\begin{equation}\label{eq:D_bounded}
    \bar{D}_\gamma(z)\approx
    \begin{cases}
        B_4+B_5z\, \ln z, & z \to 0\,, \\[1mm]
        \blue{B_6\,z^{-(\gamma+1)} \, e^{-\frac{z}{\sqrt{T}}},} & z \to \infty\,.
    \end{cases}
\end{equation}
where $B_4, B_5$ and $B_6$ are computable positive constants. We can now compare our results with those obtained by sampling $N$ IID variables from the steady-state average density $\rho_N(x)$ in Eq. (\ref{rho_weib}), which in this case belongs to the Gumbel class (cf. Appendix~\ref{app:iidvar}).
It means that the scaling function $\tilde{D}(z)$ of the distribution of $d_1$ in the i.i.d. case is given by
\begin{equation}\label{eq:Dtilde_bounded}
    \tilde{D}(z)= e^{-z} \quad \text{for} \quad z\geq 0 \;.
\end{equation}

\blue{We see that \eqref{eq:D_bounded} has a power law correction to the large $z$ tail with respect to \eqref{eq:Dtilde_bounded}.}
The small $z$ behavior of \eqref{eq:D_bounded} instead is completely different from that of \eqref{eq:Dtilde_bounded}, but it coincides with that found in the two previous cases with $\psi(\tau)$ an exponential or a power-law distribution. 

\begin{figure}[t]
\centering
\includegraphics[width=0.9\linewidth]{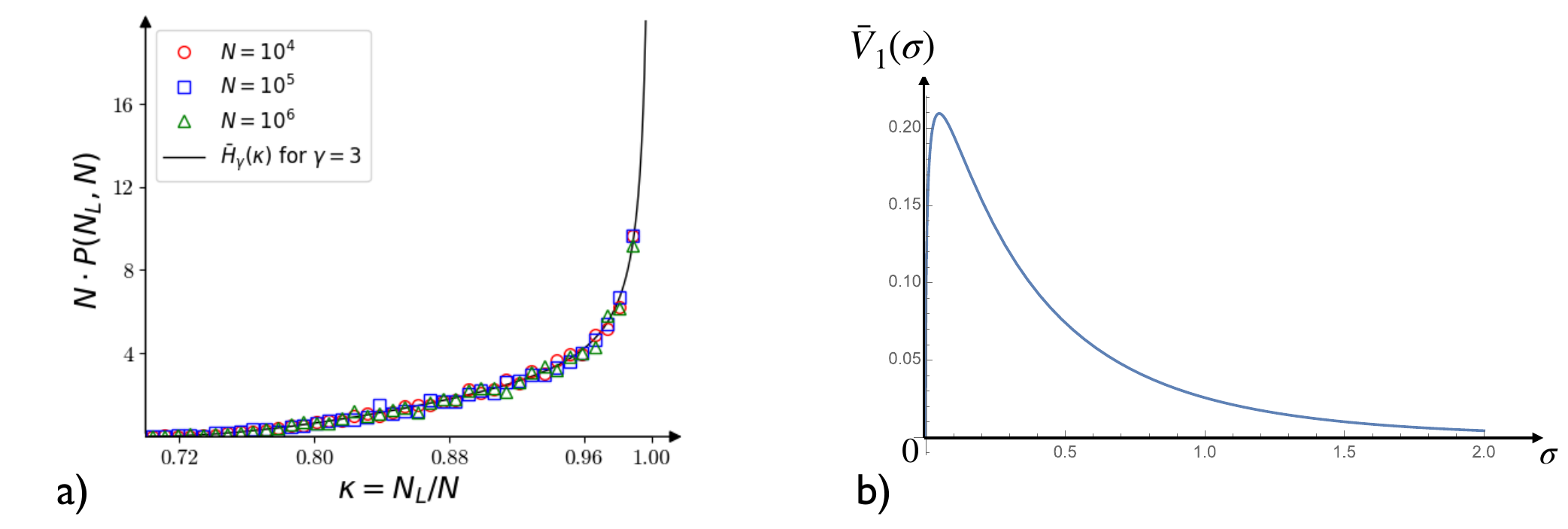}
\caption{{\bf a)}: Plot of the full counting statistics given by the distribution $P(N_L,N)\approx \frac{1}{N}\bar{H}_\gamma\left( \frac{N_L}{N} \right)$ for large $N$. The scaling function $\bar{H}_\gamma(\kappa)$ is given by \eqref{eq:H_weib_explicit}. The solid black curve shows the theoretical prediction for the resetting protocol $\psi(\tau)=\frac{\gamma r}{T^{\gamma}} \left(T-r\tau \right)^{\gamma-1}$ for \(\tau \in[0,T/r]\), $\gamma=3$ and $T=1$. The symbols indicate simulation results for different values of $N$. The simulations were done with $r=1$ and $D=\frac{1}{2}$. {\bf b)}: Plot of the function $\bar{V}_1(\sigma)$ vs $\sigma$, as given in Eq. (\ref{Vgamma}), setting $\gamma=1$.}\label{FCS_Weib}
\end{figure}

\vspace*{0.5cm}
\noindent{\it $\bullet$ Full counting statistics and variance.}
The FCS also has the same large $N$ scaling as in Eq. (\ref{eq:fullcount_tot}) with the scaling function $\bar{H}_\gamma(\kappa)$ given by
\blue{\begin{eqnarray}\label{eq:H_weib_explicit}
    \bar{H}_\gamma(\kappa)=
    \begin{cases}
        0, & 0<\kappa<\operatorname{erf}\!\left( \sqrt{\frac{\sigma}{T}} \right)\,, \\
        &\\
        \displaystyle \sigma\, \sqrt{\pi}\,\frac{\gamma+1}{T^{\gamma+1}} \left(T-\frac{\sigma}{u(\kappa)^2} \right)^{\gamma} u(\kappa)^{-3} \exp\!\left(u(\kappa)^2\right), & \operatorname{erf}\!\left( \sqrt{\frac{\sigma}{T}} \right)< \kappa < 1\,, \\[2mm]     
    \end{cases}
\end{eqnarray}}

where \(u(\kappa)=\operatorname{erf}^{-1}(\kappa)\) and \(\sigma=\frac{rL^2}{4D}\). As \(\kappa \to 1^-\), the behavior is

\blue{\bea
\bar{H}_\gamma(\kappa)\approx \frac{\sigma\, \left( \frac{\gamma+1}{T}\right)}{(1-\kappa)\, \left|\ln (1-\kappa)\right|^2},
\eea}

which diverges in an integrable manner, exactly as in the Poissonian and power-law cases. In contrast, at the lower edge of the support when $\kappa \to \operatorname{erf}\!\left( \sqrt{\frac{\sigma}{T}} \right)$, one finds that 
as \(u(\kappa)\to \sqrt{\frac{\sigma}{T}}\) and, consequently, the scaling function vanishes as a power law with a continuously varying  exponent $\gamma$ \blue{(see Appendix \ref{app:appendix_calcoli})}:

\blue{\begin{equation}\label{eq:Hgammasmallk}
    \bar{H}_\gamma(\kappa)\approx(\gamma+1)\, \pi^{\frac{\gamma+1}{2}} \, \left( \frac{T}{\sigma}\right)^{\frac{1+\gamma}{2}} e^{\frac{\sigma}{T}(1+\gamma)} \left(\kappa-\text{erf}\left(\sqrt{\frac{\sigma}{T}}\right)\right)^{\gamma}.
\end{equation}}

Interestingly, the fact that $\bar{H}_\gamma(\kappa)$ is supported over $\kappa \in [\operatorname{erf}\!\left( \sqrt{\frac{\sigma}{T}} \right),1]$
indicates that, for a fixed box size \(L\), at least a fraction $\kappa_c=\frac{N_L}{N}=\operatorname{erf}\!\left(\sqrt{\frac{rL^2}{4DT}}\right)$ of the particles is always contained within the box with probability one in the large $N$ limit. Fig.~\ref{FCS_Weib} a) shows a comparison between the theoretical prediction and simulation data for different values of $N$, confirming the expected behavior of $\bar{H}_\gamma(\kappa)$.

We now turn to the variance of $N_L$. Substituting $h(\tau)$ from Eq. (\ref{def_h_weib}) in the general formula \eqref{def_Var2} we find that the variance of $N_L$ can be written under the scaling form
\bea \label{Var_weib}
{\rm Var}(N_L) = N\, \bar{V}_\gamma\left( \sigma = \frac{r\, L^2}{4 D\,T}\right) \;,
\eea
where the scaling function $\bar{V}_\gamma(\sigma)$ is given by
\bea \label{Vgamma}
\bar{V}_\gamma(\sigma) = 2 (\gamma +1)\, \sigma\, \int_{\sqrt{\sigma}}^\infty du\, u^{-2\gamma-3}\, (u^2-\sigma)^\gamma {\rm erf}(u) {\rm erfc}(u) \;.
\eea
The scaling function $V_\gamma(\sigma)$ is again non-monotonic as a function of $\sigma$ with asymptotic behaviors given by 
\bea \label{asympt_V_beta}
\bar{V}_\gamma(\sigma) \approx
\begin{cases}
& 2 \dfrac{\Gamma(2+\gamma)}{\Gamma(3/2+\gamma)}\,\sqrt{\sigma} \quad, \quad \quad \, \;\, \sigma \to 0 \\
& \\
& \dfrac{\Gamma(\gamma+2)}{\sqrt{\pi}\, \sigma^{\gamma+3/2}}\, e^{-\sigma} \quad, \quad \quad \quad \; \sigma \to \infty \;,
\end{cases}
\eea 
In Fig. \ref{FCS_Weib} b) we show a plot of this function $\bar{V}_\gamma(\sigma)$ for $\gamma =1$, demonstrating the non-monotonic behavior.

\section{Observables for a general resetting protocol}\label{sec:risultati}

Following the examples illustrated in the previous section~\ref{sec:three_examples}, we can now present the results for the observables for a general resetting protocol characterized by the inter-reset distribution $\psi(\tau)$, or equivalently in terms of the effective distribution \( h(\tau) \). 
\bea \label{h_gen}
h(\tau) = \frac{\Psi(\tau)}{\int_0^\infty d\tau\, \Psi(\tau)} \quad, \quad {\rm where} \quad \Psi(\tau) = \int_\tau^\infty \psi(t)\,dt \;.
\eea
Our main finding is that the scaling function of each observable we have considered is entirely determined by the resetting protocol, which is encoded in the function \( h(\tau) \). We identify certain universal features that are independent of the the specific form of \( h(\tau) \). However, there are also some partially universal features that depend only on the tail behavior of the function $h(\tau)$ for large $\tau$.

To discuss the results for a general resetting protocol $\psi(\tau)$, we restrict ourselves to cases where $\langle \tau \rangle  =\int_0^\infty d\tau\, \tau \psi(\tau) = \int_0^\infty \Psi(\tau)\,d\tau$ is finite. This is also the condition for the stationary state to exist. In this case, the effective distribution $h(\tau)$ in Eq. (\ref{h_gen}) can be expressed as
\bea\label{htilde}
h(\tau)\,d\tau = \frac{1}{\langle \tau \rangle}\,\Psi(\tau) \, d\tau \;.
\eea
Therefore we can identify a time scale $\langle \tau \rangle$ and an effective resetting rate $r = 1/\langle \tau \rangle$. Consequently, $h(\tau)$ can be expressed in a dimensionless form 
\begin{equation}\label{eq:htildedef}
    h(\tau)=r\,\tilde{h}(r\tau) \quad, \quad {\rm where} \quad r = \frac{1}{\langle \tau \rangle} \;.
\end{equation}
For a general $h(\tau)$, the stationary JPDF ${\cal P}^*(\vec x)$ in Eq. (\ref{eq:SS_ciid_conh}) has a CIID structure, as explained in Section \ref{sec:NESS}. This CIID structure allows us to express the leading large $N$ behavior of different observables in terms of the single function $h(\tau)$. We show below that the scaling functions of most of these observables can be expressed in terms of the dimensionless protocol $\tilde h(y)$. We first note that there is a natural length scale in the problem
\bea \label{l_gen}
      l \;=\; \sqrt{\frac{D}{r}} \quad, \quad {\rm where} \quad r = \frac{1}{\langle \tau \rangle} \;,
\eea
that denotes the typical distance travelled by a single walker between two consecutive resettings. Thus many global observables, such as the average density, can be expressed in a dimensionless scaling form, once the distance $x$ is scaled by the microscopic length $l$.

One can also consider local observables, such as the order statistics, denoting the distribution of the $k$-th maximum $M_k$ counted from the right. The typical scale of the fluctuations of $M_k$ is set by another length scale (as seen in the examples discussed in the previous section)
 \bea \label{Lambda_gen}
      \Lambda(\alpha)
      \;=\; \sqrt{\frac{4D}{r}}\;\mathrm{erfc}^{-1}(2\alpha) \quad, \quad {\rm with} \quad r = \frac{1}{\langle \tau \rangle} \;,
 \eea
 where \(\alpha=\frac{k}{N}\). Similarly the typical spacing between the $k$-th and the $(k+1)$-th maxima, with $k = \alpha\, N$, is set by  
\bea \label{lambda_gen}
      \lambda_N(\alpha)
      \;=\;\frac{1}{N\,b\,\sqrt{r}}
      \quad\text{with}\quad
      b \;=\;\frac{\exp\!\bigl(-[\mathrm{erfc}^{-1}(2\alpha)]^2\bigr)}{\sqrt{4\pi D}} \;.
\eea

This means that most particles are confined within a region of size \(2l\) about the origin, over which the density decays. In the bulk the particles are very dense, with a typical spacing \(\lambda_N(\alpha)\sim O(1/N)\), whereas at the edges the spacing becomes much larger \(\lambda_N(\alpha)\sim O(1/\sqrt{\ln N})\). Typically, the particles in the bulk are located at distances \(\Lambda(\alpha)\sim O(1)\) from the origin, while the rightmost particle is at a distance of order \(\sqrt{\ln N}\). 
The dependence on the resetting protocol is instead entirely encoded in the scaling functions through the distribution $\tilde{h}(z)$. In fact, the scaling functions for the density $\rho_N(x)$, for the order statistics \(\text{Prob}[M_k = \omega]\), for the gap statistics \(\text{Prob}[d_k = g]\), and for the full counting statistics \(P(N_L,N)\) are given respectively by:
\begin{eqnarray}
&& \rho_N(x) \approx \frac{1}{l}\, R\left( \frac{x}{l}\right) \quad, \quad {\rm where} \quad   R(z) = \frac{1}{\sqrt{\pi}}\int_0^\infty dy\, \tilde{h}\bigl( y^2\bigr) \exp\!\left(-\frac{z^2}{4y^2}\right), \label{eq:R} \\[1mm]
&& \text{Prob}[M_{k = \alpha N} = \omega] \approx \frac{1}{\Lambda(\alpha)}  S\left(\frac{\omega}{\Lambda(\alpha)}\right) \quad, \quad {\rm where} \quad   S(z) = 2z\, \tilde{h}\bigl(z^2\bigr), \label{eq:S} \\[1mm]
&&   \text{Prob}[d_{k = \alpha N} = g] \approx \frac{1}{\lambda_N(\alpha)} D\left( \frac{g}{\lambda_N}\right) \quad, \quad {\rm where} \quad  
D(z) = 2\int_{0}^{\infty} dy\, \tilde{h}\bigl(y^2\bigr) \, \exp\!\left(-\frac{z}{y}\right), \label{eq:D} \\[1mm] 
&& P(N_L,N) \approx \frac{1}{N_L} H\left( \frac{N_L}{N}\right) \quad, \quad {\rm where} \quad H(\kappa) = \sigma\sqrt{\pi}\; u(\kappa)^{-3}\, \exp\!\left[u(\kappa)^2\right]\, \tilde{h}\!\Bigl(\frac{\sigma}{u(\kappa)^2}\Bigr), \quad \text{with} \quad \sigma = \frac{rL^2}{4D}. \label{eq:H}
\end{eqnarray}
We recall once again that $r= 1/\langle \tau \rangle$ is assumed to be finite \blue{and that $u(\kappa)=\text{erf}^{-1}(\kappa)$.} 

The first three of these scaling functions exhibit universal behavior at small values of their arguments, while \( H(k) \) shows universal behavior as \( k \to 1^- \). In fact, from the definition
\[
h(\tau=0)=\frac{\Psi(0)}{\int_0^{\infty}dt\, \Psi(t)}=\frac{1}{\langle \tau \rangle},
\]
we see that as long as the non-equilibrium steady state exists, \(h(0)\) is a positive constant, namely the effective resetting rate. As shown in the Appendix \ref{app:appendix_calcoli}, the small-\(z\) asymptotics of Eqs.~\eqref{eq:R}--\eqref{eq:D} simplify to 
\begin{align*}
    R(z) &\approx R_1^0 - R_2^0 |z|, \\
    S(z) &\approx 2\, \tilde{h}(0)\,z, \\
    D(z) &\approx D_1^0 + D_2^0\, z \ln z \;.
\end{align*}
\blue{The constants $R_1^0$, $R_2^0$, $D_1^0$ and $D_2^0$  are positive and their specific value only depend on $\tilde{h}(0)$.}
Similarly (see Appendix \ref{app:appendix_calcoli}), for the full counting statistics one obtains, as \(\kappa\to 1^{-}\),
\blue{\begin{equation}
    H(\kappa) \approx \frac{\sigma \tilde{h}(0)}{(1-\kappa) [ \ln(1-\kappa) ]^2 }.
\end{equation}}

There are also some universal behaviors at large $z$ of the scaling functions $R(z), S(z)$ and $D(z)$, depending on the large $y$ tail of the function $\tilde h(y)$. These tails of $\tilde h(y)$ turns out to be of three different types, as in the study of the EVS for IID variables \cite{AFirstCourseinOrderStatistics,StatisticsofExtremesandRecordsinRandomSequence,Extremevaluestatisticsofcorrelatedrandomvariables:Apedagogicalreview}. 
These three tail behaviors are given by
\begin{enumerate}[label=(\Roman*),ref=(\Roman*)]
    \item \label{case:exp} $\tilde{h}(y)\sim e^{-y^{\nu}}$ for large $y$ with $\nu > 0$, 
    \item \label{case:bounded} $\tilde{h}(y)\sim (T-y)^{\gamma}$ for $y \in [0,T]$ and $\tilde{h}(y)=0$ outside this domain (here $\gamma>-1$),
    \item \label{case:power} $\tilde{h}(y)\sim y^{-\beta}$ for large $y$, with $\beta > 1$.
\end{enumerate}

We have defined the three classes in this way because of the way in which their asymptotics can be treated (cf. Appendix \ref{app:appendix_calcoli}).
For example, in the scaling function of the density \eqref{eq:R}, the integral contains an exponential term. As a result, in the limit \( z \to \infty \), the integral is dominated by the large-\( y \) behavior of \( \tilde{h}(y) \) and we can then substitute the asymptotic form of \( \tilde{h}(y) \) into it. In case~\ref{case:exp}, the integral is dominated by a saddle point; in case~\ref{case:bounded}, the dominant contribution comes from the boundary of the domain at \( T \); in case~\ref{case:power}, a simple change of variables is sufficient to evaluate the asymptotics.
In this way we get (cf. Appendix~\ref{app:appendix_calcoli}):
\begin{equation}
R(z)\approx
    \begin{cases}
        R_1^{\infty} \, z^{\frac{1-\nu}{\nu+1}} \exp\left( -R_2^{\infty} \, z^{\frac{2\nu}{\nu+1}}\right) & \text{for type (I)}
        \\
        \blue{R_3^{\infty}\, z^{-2(\gamma+1)} \, e^{-\frac{z^{2}}{4T}}} & \text{for type (II)}
        \\
        \frac{R_4^{\infty}}{z^{2\beta-1}} & \text{for type (III)}
    \end{cases}
\end{equation}
where $R_i^{\infty}$ for $i=1,2,3,4$ are positive constants.
We see that in case \ref{case:exp}, i.e. when $\tilde{h}(\tau)\sim e^{-\tau^{\nu}}$ for large $\tau$, the density has also a stretched exponential behavior at large \(z\) with an exponent that lies in the interval \((0,2)\): it approaches \(0\) as \(\nu\) becomes small and tends to \(2\) for \(\nu\to \infty\). This is interesting since, as we can see,
a stretched exponential with an exponent $2$ is obtained for \ref{case:bounded}, i.e. when $\tilde{h}(\tau)$ has a domain bounded from above. This makes sense since a bounded domain $[0,T]$ means that in each time interval $T$ there is for sure at least a resetting event, which is not true when $\psi(\tau)$ is unbounded.\\

For the function $S(z)$ it is easy to see that $S(z)\approx S_1^{\infty}\,z\,e^{-z^{2\nu}} $ when $\tilde{h}(y)$ is in class \ref{case:exp}, and $S(z)\approx \frac{S_2^{\infty}}{z^{2\beta-1}}$ for class \ref{case:power}. For the case \ref{case:bounded} $S(z)$ has a bounded domain at $z= \, \sqrt{T}$. It approaches that point as $S(z)\approx S_3^{\infty}z(T-z^2)^{\gamma}$.
This implies that the probability that the rightmost particle is at distance $x$ from the origin becomes exactly 0 in the $N \to \infty$ limit for $x<0$  or for $x>L_N \, \sqrt{T}$, with $L_N = \sqrt{\frac{4D\ln N}{r}}$. Thus, at leading order, the gas is confined in a symmetric region of size $2\,L_N \, \sqrt{T}$ around the origin.\\

Regarding the scaling function of the spacing distribution, since in \eqref{eq:D} we have an exponential term inside the integral, we find again that the integral is dominated by a saddle point for \ref{case:exp}, for \ref{case:bounded} it is dominated by the boundary of the domain of $\tilde{h}(y)$ and for \ref{case:power} we just have to make a change of variable.
We obtain (cf. Appendix~\ref{app:appendix_calcoli}) respectively that, in the limit of large $z$:

\begin{equation}
D(z)\approx
    \begin{cases}
        D_1^{\infty} z^{\frac{1-\nu}{2\nu+1}} \exp\left( - D_2^{\infty}\, z^{\frac{2\nu}{2\nu+1}}\right) & \text{for type (I)}
        \\
         \blue{D_3^{\infty}\, z^{-(\gamma+1)}\, e^{-\frac{z}{\sqrt{T}}}} & \text{for type (II)}
        \\
        \frac{D_4^{\infty}}{z^{2\beta-1}} & \text{for type (III)}
    \end{cases}
\end{equation}
where $D_i^{\infty}$ for $i=1,2,3,4$ are positive constants.
As before for the density, in case \ref{case:exp} the tail is a stretched exponential with an exponent $\theta=\frac{2\nu}{2\nu+1}$ that this time lies in the interval $(0,1)$. As for the density, the upper limit $\theta=1$ is attained for $\nu \to \infty$ and in corresponds to the behavior of $D(z)$ in the case of $\tilde{h}(y)$ of type \ref{case:bounded}, i.e. with a bounded domain.
When instead we are in case \ref{case:power} we get a power-law tail.\\

 Similarly, the small $\kappa$ behavior of $H(\kappa)$ also exhibits some universalities. We recall that the scaling function of the FCS is given by
\begin{equation}\label{eq:FCS_ripetuta}
    H(\kappa) = \sigma\sqrt{\pi}\; u(\kappa)^{-3}\, \exp\!\left[u(\kappa)^2\right]\, \tilde{h}(y^*)\,,
\end{equation}
where
\bea \label{u_Hk}
u(\kappa)=\operatorname{erf}^{-1}(\kappa),\quad \sigma=\frac{rL^2}{4D},\quad \quad y^*= \frac{\sigma}{\,u(\kappa)^2} \quad {\rm and} \quad r = \frac{1}{\langle \tau \rangle}\,.
\eea
It is evident that, given the dependence of $y^*$ on $\kappa$, the behavior of \blue{$\tilde{h}(y)$ for large $y$ } determines that of $H(\kappa)$ for small $\kappa$. 
\blue{In the Appendix~\ref{app:appendix_calcoli} we show that in this limit, when $\tilde{h}(y)$ is unbounded for large $y$, we obtain:
\begin{equation}\label{eq:FCS_kto0}
    H(\kappa)\approx \frac{8 \sigma}{\pi k^3}\, \tilde{h}\left( \frac{4 \sigma}{\pi \, \kappa^2}\right).
\end{equation}}
\blue{We can substitute the tail of $\tilde{h}(y)$ in \eqref{eq:FCS_kto0} to get:}
\blue{\begin{align}
    H(\kappa)&\approx \frac{H_1^{\infty}}{\kappa^3}\exp{\left( -\frac{\left( \frac{4\sigma}{\pi} \right)^{\nu}}{\kappa^{2\nu}}\right)} \quad \text{when $\tilde{h}(y)$ is of type \ref{case:exp}}\;,\\
    H(\kappa)&\approx H_2^{\infty} \, \kappa^{2\beta{-3}}\quad \text{when $\tilde{h}(y)$ is of type \ref{case:power}}\;.
\end{align}
The values $H_i^{\infty}$ for $i=1,2$ are positive constants.
}
\blue{Instead, for \ref{case:bounded}, where $\kappa$ is bounded from below by $\kappa_c = \operatorname{erf}\left(\sqrt{\frac{\sigma}{T}}\right)$, (see Appendix~\ref{app:appendix_calcoli}) we obtain:
\begin{equation}
    \bar{H}_\gamma(\kappa)\approx(\gamma+1)\, \pi^{\frac{\gamma+1}{2}} \, \left( \frac{T}{\sigma}\right)^{\frac{1+\gamma}{2}} e^{\frac{\sigma}{T}(1+\gamma)} \left(\kappa-\text{erf}\left(\sqrt{\frac{\sigma}{T}}\right)\right)^{\gamma}.
\end{equation}}

The case \ref{case:bounded} is particularly interesting since it tells us that, fixed the size of the box $L$, it contains for sure at least a fraction $\kappa_c=\frac{N_L}{N}=\text{erf}\left(\sqrt{\frac{rL^2}{4DT}}\right) $ of particles. This reflects the fact that imposing a maximum inter-reset time confines the particles more tightly around the origin than in the other cases.

Finally, we turn to the variance of $N_L$ for an arbitrary resetting protocol $h(\tau)$, given by the exact formula in Eq.~(\ref{def_Var2}). For small $L$, replacing ${\rm erf}(x) \approx (2/\sqrt{\pi})\, x$ for small $x$ and ${\rm erfc}(x) \approx 1$ one finds that the variance is generically proportional to the system size $L$ with
\bea \label{varNL_small}
{\rm Var}(N_L) \approx N \left[\frac{1}{\sqrt{\pi D}}\, \int_0^\infty d\tau\, \frac{h(\tau)}{\sqrt{\tau}} \right]\, L \quad, \quad L \to 0 \;,
\eea
provided the integral inside the brackets exists. In contrast, for large $L$, using the asymptotic behaviors ${\rm erfc}(x) \approx e^{-x^2}/(x\sqrt{\pi})$ for large $x$ and ${\rm erf}(x) \approx 1$, one finds
\bea \label{varNL_large}
{\rm Var}(N_L) \approx \frac{N\, \sqrt{4 D}}{\sqrt{\pi} L} \int_0^\infty h(\tau)\, \sqrt{\tau}\, e^{-L^2/(4 D \tau)}  \;.
\eea 
Clearly the dominant contribution to this integral comes from the large $\tau$ region and depends on the tail of $h(\tau)$ for large $\tau$. So, while the small $L$ behavior of the variance of $N_L$ is universal, the large $L$ behavior is not. In any case, it decays for large $L$, indicating a hyper uniform gas.

\section{Conclusion}\label{sec:concl}

In this work, we have introduced and analyzed a general class of non-Poissonian resetting protocols for a gas of \( N \) independent Brownian particles in one dimension, extending the classical Poissonian case, characterized by the inter-reset time distribution \( \psi(\tau) = re^{-r\tau} \), to arbitrary distributions \( \psi(\tau) \). While our analysis focused on the one-dimensional case, the approach and main results are easily generalizable to higher-dimensional systems. In this model, the particles are {\it noninteracting} between resettings. However, as time grows, the simultaneous resetting introduces correlations between particles that grow with time and eventually leads to a strongly correlated (attractive all-to-all interactions) gas in a nonequilibrium stationary state (NESS).

By exploiting the renewal structure of the dynamics we have shown that the joint distribution of the particle positions in the NESS has a 
conditionally independent and identically distributed (CIID) structure. This solvable structure 
enabled the computation of a broad range of observables, including the single-particle density, extreme value statistics, order statistics, gap statistics, and full counting statistics, by expressing them as averages over diffusive processes conditioned on a fixed elapsed time since the last reset. These averages are taken over the distribution \( h(\tau) = r\tilde{h}(r\tau) \), which represents the normalized probability density of a time interval \( \tau \) without a reset.
While the detailed forms of these observables depend on the specific shape of \( \tilde{h}(z) \), their characteristic scales are universal, determined solely by the diffusion coefficient \( D \) and the typical reset rate \( r \).
We have identified three universality classes based on the tail behavior of \( \tilde{h}(z) \): stretched-exponential, power-law, and bounded support. Each class leads to distinct asymptotic forms for the scaling functions associated with the observables studied, namely \( R(z) \), \( S(z) \), \( D(z) \), and \( H(k) \), which describe the single-particle density, order and extreme value statistics, gap statistics, and full counting statistics, respectively. Beyond these class-dependent behaviors, we have also found fully universal features in each scaling function, governed solely by the value of \( \tilde{h}(0) \).
Our theoretical predictions were illustrated and validated numerically through two explicit examples, alongside the classical Poissonian case, each representing one of the universality classes. 

Our work provides a nice example of stochastic control of non-equilibrium stationary states (NESS) in the sense that, by tuning the resetting interval distribution $\psi(\tau)$, we can generate a whole class of tunable NESS that have, in addition, an exactly solvable CIID structure. This framework illustrates a mechanism to build a strongly correlated NESS out of noninteracting particles. In this sense, the correlations in the NESS in this class of models are emergent and not directly built into an energy function, as in equilibrium systems.  
An added bonus of such a NESS is its CIID structure that 
enables exact computations of several physical observables, despite the fact that the NESS is strongly correlated.   
This is particularly valuable given that strongly correlated non-equilibrium systems are in general poorly understood, and it is often not even possible to compute explicitly their steady states. Having a class of correlated models where the steady state is explicitly accessible, and where a wide range of physically measurable observables can be exactly computed, offers an important opportunity to gain insight into the behavior of out-of-equilibrium systems. Future directions include extending this approach to other many-body systems, including interacting ones (see for instance the recent study of resetting Dyson Brownian motions~\cite{biroli2025resetting}).

\blue{Recently, stochastic resetting has been experimentally realized with colloidal particles diffusing in water and being reset via laser-generated optical traps. Single-particle resetting has been implemented in several experimental works \cite{Besga_2020,Experimental_3,Experimental_1}, and only very recently these experimental studies have been extended to multi-particle systems \cite{Many-BodyColloidalDynamics...,biroli2025exp}. In particular, in \cite{biroli2025exp} four particles were held in distinct optical traps whose stiffness is synchronously modulated at random times, thereby implementing simultaneous resetting events. This setup would also allow a straightforward test of our prediction for non-Poissonian resetting protocols: one would simply need to draw the inter-reset intervals of the traps from the desired distribution $\psi(\tau)$ at each event.}
Moreover, since resetting can be applied to generic many-body stochastic processes \cite{SR_review}, our method might be useful to study other interacting many-body systems, such as the resetting Ising model~\cite{magoni2020ising}, fluctuating interfaces \cite{gupta2014fluctuating}, hard-core lattice gases \cite{basu2019symmetric}, etc.

\blue{Since all the non-trivial correlations in the system stem from the fact that all particles are reset simultaneously, it is natural to ask what would happen if, at each resetting event, only a subset of $m$ particles out of the total $N$ were reset. However, introducing this mechanism breaks the renewal structure of the process: after each resetting event, the system retains some memory of its previous state, and the dynamics can no longer be described by a standard renewal equation.
}

Finally, another interesting avenue is to explore the first-passage properties of such multi-particle systems with simultaneous resetting subject to a 
non-Poissonian resetting protocol. For the Poissonian case, it was shown that the mean first-passage time to a target, located at a position $L$, 
has an interesting transition as a function of $N$: the optimal resetting rate $r^*$ that minimises the mean first-passage time to a target is nonzero for $N < 7$, while it vanishes for $N \geq 7$ \cite{Criticalnumberofwalkers}. It would be interesting to see whether such a transition exists for non-Poissonian resetting protocols. Another interesting problem is to study the survival probability of this multi-particle system with resetting in the presence of a single absorbing center at position $L$. For a single particle, this was studied in Ref. \cite{whitehouse2013effect} where the survival probability was computed as a function of the absorption rate $a$. It would be interesting to see how this survival probability depends on the number of walkers $N$ with simultaneous resetting that makes the particles more and more correlated as time increases.

\vspace*{0.5cm}
\noindent{\it Acknowledgments.--}{We acknowledge support from ANR Grant No. ANR-23-CE30-0020-01 EDIPS. SNM and GS acknowledge hospitality of the MATRIX Institute (University of Melbourne, Creswick) during the conference {\it Log-Gases in Caeli Australi}, where this work was completed}.

\appendix

\section{EVS for IID variables}\label{app:iidvar}
We briefly summarize some classical results concerning the statistics of extremes for independent and identically distributed IID random variables. Let \( \{x_1, x_2, \ldots, x_N\} \) be IID random variables drawn from a probability density function (PDF) \( p(x) \), and let their ordered values be denoted as \( \{M_1 > M_2 > \ldots > M_N\} \). This problem is well studied in the literature~\cite{AFirstCourseinOrderStatistics,OrderStatistics,StatisticsofExtremesandRecordsinRandomSequence,Extremevaluestatisticsofcorrelatedrandomvariables:Apedagogicalreview}.

According to the Fisher--Tippett--Gnedenko theorem, in the limit \( N \to \infty \), the distribution \( \text{Prob}\{M_1 = \omega\} \), when properly centered and rescaled, converges to one of three universal forms
\begin{equation}
    \text{Prob}\{M_1 = a_N + b_N z\} \xrightarrow{N \to \infty} f_{\rho}(z) \;,
\end{equation}
where $\rho=I,II,III$ indicates the three classes. The scaling factors \( a_N \) and \( b_N \) depend on the parent distribution \( p(x) \), while the scaling function \( f_{\rho}(z) \) is universal, determined solely by the tail behavior of \( p(x) \). The three universality classes are
\begin{enumerate}
    \item \textbf{Gumbel (Class I):} When \( p(x) \) decays faster than any power law on an unbounded domain, the limiting form is the Gumbel distribution:
    \bea f_I(z) = e^{-z - e^{-z}} \;. \eea
    \item \textbf{Fr\'echet (Class II):} When \( p(x) \sim x^{-1 - \alpha} \) for large \( x \) (heavy-tailed), with $\alpha>0$, the limiting distribution is
    \bea f_{II}(z) = \frac{\alpha}{z^{1 + \alpha}} e^{-z^{-\alpha}}, \quad z > 0 \;. \eea
    \item \textbf{Weibull (Class III):} When \( p(x) \) has bounded support, i.e., \( p(x) \sim B(a - x)^{\gamma - 1} \) for \( x \leq a \), the distribution becomes:
    \bea f_{III}(z) = \gamma |z|^{\gamma - 1} e^{-|z|^{\gamma}}, \quad z < 0 \;. \eea
\end{enumerate}

For \( M_k \) with \( k = O(1) \), a similar result holds: the scaling function of the rescaled variable \( z = \frac{M_k - a_N}{b_N} \) depends on \( k \), but \( a_N \) and \( b_N \) remain the same as for \( k = 1 \). Likewise, the statistics of the gaps \( d_k = M_k - M_{k+1} \) obey the scaling form ~\cite{AFirstCourseinOrderStatistics,OrderStatistics,StatisticsofExtremesandRecordsinRandomSequence,Extremevaluestatisticsofcorrelatedrandomvariables:Apedagogicalreview}:
\bea
    \text{Prob}\{d_k = b_N z\} \xrightarrow{N \to \infty} \tilde{p}_{k, \rho}(z),
\eea
with the following asymptotic distributions
\begin{enumerate}
    \item \textbf{Gumbel (Class I):}
    \[ \tilde{p}_{k, I}(z) = \Theta(z) \, k e^{-kz} \;. \]
    \item \textbf{Fr\'echet (Class II):}
    \[ \tilde{p}_{k, II}(z) = \Theta(z) \, \frac{\alpha^2}{(k-1)!} \int_0^\infty e^{-x^{-\alpha}} x^{-\alpha - 1} (x + z)^{-\alpha k - 1} \, dx \;. \]
    \item \textbf{Weibull (Class III):}
    \[ \tilde{p}_{k, III}(z) = \Theta(z) \, \frac{\gamma^2}{(k-1)!} \int_0^\infty (x + z)^{\gamma - 1} e^{-(x + z)^{\gamma}} x^{\gamma k - 1} \, dx \;. \]
\end{enumerate}
These results are valid for \( k \geq 1 \), with \( k = O(1) \).

\section{Derivations of some analytical results}\label{app:appendix_calcoli}

In this subsection, we derive general results for the scaling functions of the observables described in Section \ref{sec:risultati} in the main text.

\subsection{Small $z$ behavior for $R(z),\, D(z),\, S(z)$}
We start by recalling that, as long as the non-equilibrium steady state exists, \( h(0) \) is a positive constant.

\subsubsection{Density profile $R(z)$}

The density of particles can be written in the scaling form
\begin{equation}
    \rho_N(x) = \frac{1}{l} R\left( \frac{x}{l} \right),
\end{equation}
where \( l = \sqrt{\frac{D}{r}} \), and
\begin{equation} \label{eq:Rz_appendix}
    R(z) = \frac{1}{\sqrt{\pi}} \int_0^\infty \mathrm{d}u\, \tilde{h}(u^2) \exp\left(-\frac{z^2}{4u^2}\right).
\end{equation}

We aim to show that, since $\tilde{h}(0)$ is a non-zero value, the small-\( z \) behavior of \( R(z) \) is
\bea
R(z) \approx R_1^0 - R_2^0 |z| \;.
\eea
Instead of analysing $R(z)$ in Eq. \eqref{eq:Rz_appendix}, it is more convenient to study its derivative
\begin{equation}
    R'(z) = \frac{dR(z)}{dz} = -\frac{z}{2\sqrt{\pi}} \int_0^\infty \mathrm{d}u\, \frac{\tilde{h}(u^2)}{u^2} \exp\left(-\frac{z^2}{4u^2}\right).
\end{equation}
Performing the change of variables \( y = z/u \), we obtain
\begin{equation}
    R'(z) = -\frac{1}{2\sqrt{\pi}} \int_0^\infty \mathrm{d}y\, \tilde{h}\left( \frac{z^2}{y^2} \right) \exp\left(-\frac{y^2}{4} \right).
\end{equation}
In the limit \( z \to 0 \), the argument of \( \tilde{h} \) tends to zero, so we can approximate \( \tilde{h}(z^2/y^2) \approx \tilde{h}(0) \). Thus,
\[
R'(z) \xrightarrow[]{z \to 0} -\frac{\tilde{h}(0)}{2\sqrt{\pi}} \int_0^\infty \mathrm{d}y\, \exp\left(-\frac{y^2}{4} \right),
\]
which is a finite constant. By integrating over $z$, this implies that \( R(z) \approx R_1^0 - R_2^0 |z| \) for small \( z \), consistently with the even symmetry of the density profile. The constants are given by
\begin{equation}
    R_1^0 = R(0) = \frac{1}{\sqrt{\pi}} \int_0^\infty \mathrm{d}u\, \tilde{h}(u^2), \qquad
    \blue{R_2^0 = \frac{ \tilde{h}(0)}{2\sqrt{\pi}} \int_0^\infty \mathrm{d}y\, \exp\left(-\frac{y^2}{4} \right)=\frac{\tilde{h}(0)}{2} .}
\end{equation}

\subsubsection{Spacing distribution $D(z)$}

We now consider the scaling function associated with the spacing distribution
\begin{equation}
    D(z) = 2 \int_0^\infty \mathrm{d}u\, \tilde{h}(u^2) \exp\left( -\frac{z}{u} \right) \;.
\end{equation}
As done above for $R(z)$, we consider the derivative $D'(z)$ and perform the change of variable $y=\frac{z}{u}$ inside the integral. This leads to
\begin{equation}
    D'(z)\approx-2\int_z^{\infty}dy \, \frac{e^{-y}}{y}\, \tilde{h}\left( \frac{z^2}{y^2}\right) \;.
\end{equation}
As $z\to0$ we can set $\tilde{h}\left( \frac{z^2}{y^2}\right)=\tilde{h}(0)$ and perform an integration by parts. This leads to $D'(z)\propto \ln z$ and, after one more integration, one finds 
\bea
D(z) \approx D_1^0 + D_2^0\, z \ln z \;,
\eea
where the constants $D_1^0$ and $D_2^0$ can in principle be straightforwardly computed. 

\subsubsection{Order and extreme value statistics $S(z)$}
In this case the scaling function is $S(z) = 2z\, \tilde{h}\bigl(z^2\bigr) \label{eq:S}$. This trivially gives $S(z) \xrightarrow[]{z\to0} 2\tilde{h}\bigl(0\bigr)z $, which is a linear behavior independently of the resetting protocol.

\subsection{Large $z$ behavior for $R(z),\, D(z),\, S(z)$}

We now study the large-\( z \) behavior of $R(z),\, D(z),\, S(z)$ for the three universality classes defined as
\begin{enumerate}[label=(\Roman*),ref=(\Roman*)]
    \item $\tilde{h}(y)\sim e^{-y^{\nu}}$ for large $y$.
    \item $\tilde{h}(y)\sim (T-y)^{\gamma}$, for $y \in [0,T]$ and $\tilde{h}(y)=0$ outside this domain.
    \item $\tilde{h}(y)\sim y^{-\beta}$ for large $y$.
\end{enumerate}
For the first case \ref{case:exp}, in the limit \( z \to \infty \), the integral in \eqref{eq:Rz_appendix} is dominated by large \( u \). Substituting the asymptotic form, we write
\bea \label{R_sp}
R(z) \approx \frac{1}{\sqrt{\pi}} \int_0^\infty \mathrm{d}u\, \exp\left[-f(u)\right], \quad \text{with} \quad f(u) = u^{2\nu} + \frac{z^2}{4u^2} \;.
\eea
To apply the saddle point method we notice that the minimum of \( f(u) \) occurs at
\bea
u^* = \left( \frac{z^2}{4\nu} \right)^{1/(2\nu + 2)}.
\eea
Expanding $f(u)$ in Eq. (\ref{R_sp}) up to the second order around $u^*$ and then performing a Gaussian integral gives the leading behavior
\begin{equation}
    R(z) \approx R_1^{\infty}\, z^{\frac{1-\nu}{\nu+1}} \exp\left( -R_2^{\infty}\, z^{\frac{2\nu}{\nu + 1}} \right),
\end{equation}
where \( R_1^{\infty} \) and \( R_2^{\infty} \) are positive constants which depends on \( \nu \).

Using the same procedure, we find that for large \( z \),
\begin{equation}
    D(z) \approx D_1^{\infty}\, z^{\frac{1 - \nu}{2\nu + 1}} \exp\left( - D_2^{\infty}\, z^{\frac{2\nu}{2\nu + 1}} \right),
\end{equation}
with positive constants \( D_1^{\infty} \) and \( D_2^{\infty} \).\\

When considering instead the case \ref{case:bounded}, the integrals defining $R(z)$ and $D(z)$ are given by
\begin{align}
    R(z) &= \frac{1}{\sqrt{\pi}} \int_0^{\sqrt{T}}\mathrm{d}u\, \tilde{h}(u^2) \exp\left(-\frac{z^2}{4u^2}\right) \;, \\
    D(z) &= 2 \int_0^{\sqrt{T}} \mathrm{d}u\, \tilde{h}(u^2) \exp\left( -\frac{z}{u} \right) \;.
\end{align}
For large $z$ they are dominated by the boundary $u\sim \sqrt{T}$. This suggests that we can substitute the tail \ref{case:bounded} of $\tilde{h}(y)$ and perform the change of variable $u= \sqrt{T}(1-\epsilon)$. This leads to
\begin{align}
    R(z) &= \sqrt{\frac{T}{\pi}} \int_0^1 \mathrm{d}\epsilon\, \left( T-T(1-\epsilon)^2\right)^{\gamma} \exp\left(-\frac{z^2}{4T(1-\epsilon)^2}\right)\\
    D(z) &= 2 \sqrt{T}\int_0^1 \mathrm{d}\epsilon\, \left( T-T(1-\epsilon)^2\right)^{\gamma} \exp\left( -\frac{z}{\sqrt{T}(1-\epsilon)} \right)
\end{align}
For $z\to \infty$ only the values $\epsilon\sim0$ will matter and we can further approximate $u^2=T(1-\epsilon)^2\approx T(1-2\epsilon)$, $1/u^2\approx \frac{1+2\epsilon}{T}$ and $1/u \approx \frac{1+\epsilon}{\sqrt{T}}$. We obtain

\blue{\begin{align}
    R(z) &= \sqrt{\frac{T}{\pi}}(2T)^{\gamma} e^{-\frac{z^2}{4T}} \int_0^1 \mathrm{d}\epsilon\, \epsilon^{\gamma} \exp\left(-\frac{z^2}{2T}\epsilon\right)=\sqrt{\frac{T}{\pi}}\, (2T)^{2\gamma+1}\, \Gamma(\gamma+1)\, z^{-2(\gamma+1)}\, e^{-\frac{z^2}{4T}} \;, \\
    D(z) &= 2 \sqrt{T}(2T)^{\gamma} e^{-\frac{z}{\sqrt{T}}}\int_0^1 \mathrm{d}\epsilon\, \epsilon^{\gamma} \exp\left( -\frac{z}{\sqrt{T}} \epsilon \right)=2^{\gamma+1}\, T^{\frac{3\gamma+2}{2}}\,\Gamma(\gamma+1) \, z^{-(\gamma+1)} \, e^{-\frac{z}{\sqrt{T}}}\, \;,
\end{align}}
where we have used 
\blue{\begin{equation}
    \lim_{a\to \infty} \int_0^1 x^{\gamma} e^{-a\,x} \, dx = \frac{\Gamma(\gamma+1)}{a^{\gamma+1}} 
\end{equation}
where $\Gamma(s)=\int_0^{\infty} t^{\,s-1} e^{-t} \, dt$ is the Gamma function.}\\

We now consider class \ref{case:power}, i.e. $\tilde{h}(y)\sim y^{-\beta}$ when $y$ is large. When $z\to \infty$ the scaling functions $R(z)$ and $D(z)$ are again dominated by large values of $u$ in the integration domain. We can then substitute the tail of $\tilde{h}$ in \eqref{eq:R} and \eqref{eq:D} to get
\begin{align}
    R(z) &= \frac{1}{\sqrt{\pi}} \int_0^{\infty}\mathrm{d}u\,\frac{1}{u^{2\beta}} \exp\left(-\frac{z^2}{4u^2}\right) \;, \\
    D(z) &= 2 \int_0^{\infty} \mathrm{d}u\,\frac{1}{u^{2\beta}} \exp\left( -\frac{z}{u} \right) \;.
\end{align}
Performing the change of variable $\frac{z}{u}=y$, and using the relation $\int_0^{\infty}dy\, y^{2\beta-2}e^{-y^2}=\frac{1}{2}\Gamma(\beta-\frac{1}{2})$, where $\Gamma(x)$ is the gamma function defined by $\Gamma(x)=\int_0^{\infty} dy \, y^{x-1} \, e^{-y}$, we obtain for large $z$
\begin{align}
    R(z) & \approx \frac{2^{2\beta-2}\Gamma(\beta-\frac{1}{2})}{\sqrt{\pi}} \frac{1}{z^{2\beta-1}} \;, \\
    D(z) & \approx  \frac{2\, \Gamma\left(2\beta-1\right)}{z^{2\beta-1}} \;.
\end{align}
The expansion for $R(z)$ is valid a priori as long as $\beta>\frac{1}{2}$, but we consider here only the case $\beta>1$ in order to have a stationary state.

The $z\to \infty$ behavior of $S(z)$ is instead trivial since one just needs to insert the three possible tails in the expression $S(z) = 2\, \tilde{h}(z)\,z$.

\subsection{Asymptotics for the FCS \( H(\kappa) \)}

The function \( H(\kappa) \) is defined as
\begin{equation} \label{eq:H_appendix_clean}
    H(\kappa) = \sigma \sqrt{\pi} \, u(\kappa)^{-3} \exp\left( u(\kappa)^2 \right) \tilde{h}(y^*),
\end{equation}
where \blue{we recall from equation \eqref{eq:htildedef} that $h(\tau)=r\tilde{h}(r \tau)$ where $h(\tau)$ is the effective input interval distribution in the protocol},
\( u(\kappa) = \text{erf}^{-1}(\kappa) \), and \( y^* = \sigma / u(\kappa)^2 \) with \(\sigma=\frac{rL^2}{4D}\).\\

\blue{\noindent{\bf{The limit $\kappa\to1$.}}
We first consider the $\kappa\to1$ limit.
Since $\text{erf}(u)=\kappa$, it is clear that $u\to \infty$ as $\kappa\to1$. Using the asymptotic behavior of the error function for large argument (as $u\to \infty$) one gets:
\begin{equation}\label{eq:erf(u)limit}
    1- \frac{e^{-u^2}}{u \sqrt{\pi}}\approx \kappa.
\end{equation}
This gives:
\begin{equation}\label{eq:eallau2}
    e^{u^2}\approx \frac{1}{u \sqrt{\pi} (1-\kappa)}.
\end{equation}
Substituing this behavior in \eqref{eq:H_appendix_clean} and using the fact that $y^*=\frac{\sigma}{u(\kappa)^2}\to 0$ as $\kappa \to 1$, we get to leading order 
\begin{equation}\label{eq:H(k)kto1}
    H(\kappa)=  \frac{\sigma \, \tilde{h}(0)}{(1-\kappa) \, u(\kappa)^4},
\end{equation}
where we assumed that $\tilde{h}(0)$ is non-zero.
From \eqref{eq:eallau2} we see that $u(\kappa) \approx \sqrt{-\ln(1-\kappa)}$ as $\kappa\to 1$. Substituting this behavior in \eqref{eq:H(k)kto1} gives the leading order asymptotics of $H(\kappa)$ as $\kappa\to1$:
\begin{equation}
    H(\kappa) \approx \frac{\sigma \tilde{h}(0)}{(1-\kappa) [ \ln(1-\kappa) ]^2 }.
\end{equation}}\\

\blue{\noindent{\bf{The limit $\kappa\to 0.$}} We now investigate the behavior as $\kappa\to 0$ limit. First, we consider the cases where $\tilde{h}(y)$ is unbounded for large $y$. In this situation, we can use the small $u$ expansion of $\text{erf}(u)\approx\frac{2}{\sqrt{\pi}}u$, which leads to:
\begin{equation}
   u(\kappa) \approx \frac{\sqrt{\pi}}{2} \kappa .
\end{equation}
This results in 
\begin{equation}
    y^*= \frac{\sigma}{u(\kappa)^2}\approx \frac{4 \sigma}{\pi \kappa^2}.
\end{equation}
Thus, we obtain 
\begin{equation}
    H(\kappa)\approx \frac{\sigma \sqrt{\pi}}{\left( \frac{\sqrt{\pi}}{2} \kappa \right)^3} \, \tilde{h}\left( \frac{4 \sigma}{\pi \, \kappa^2}\right)=\frac{8 \sigma}{\pi k^3}\, \tilde{h}\left( \frac{4 \sigma}{\pi \, \kappa^2}\right).
\end{equation}
Since we are in the small $\kappa$ limit, we see that the argument of the function $\tilde{h}$ diverges as $\kappa\to 0$.
Hence, this limit is non-universal and depends on the large $y$ behavior of the input protocol $\tilde{h}(y)$.
}

\blue{In the case of a bounded resetting protocol instead, $\kappa$ is constrained from below by the value $\kappa_c = \operatorname{erf}\left(\sqrt{\frac{\sigma}{T}}\right)$, meaning that $\kappa$ cannot approach zero. Therefore, for $\kappa \sim \kappa_c^+$, we can use the following approximation for $u(\kappa) = \operatorname{erf}^{-1}(\kappa)$:
\[
u(\kappa) \approx \operatorname{erf}^{-1}(\kappa_c) + \frac{\sqrt{\pi}}{2} e^{\frac{\sigma}{T}} (\kappa - \kappa_c).
\]
Since $\operatorname{erf}^{-1}(\kappa_c) = \sqrt{\frac{\sigma}{T}}$, this simplifies to:
\[
u(\kappa) \approx \sqrt{\frac{\sigma}{T}} + \frac{\sqrt{\pi}}{2} e^{\frac{\sigma}{T}} (\kappa - \kappa_c).
\]
Substituting this expression into \eqref{eq:H_weib_explicit} and retaining the leading-order term gives the desired result in \eqref{eq:Hgammasmallk}.}

\newpage

\end{document}